\documentclass[twocolumn]{aastex62}    
\usepackage{graphicx,times,epsf}             
\usepackage{longtable}
\usepackage{natbib}
\usepackage{amssymb}
\usepackage{amsmath}
\usepackage{psfig}
\usepackage{txfonts}
\usepackage[T1]{fontenc}
\usepackage{soul,xcolor}

\shortauthors{Yang et al.}
\begin{document}

\title{Revisiting KELT-19Ab, WASP-156b and WASP-121b in the TESS Era}

\email{Fan Yang: sailoryf@nao.cas.cn, sailoryf1222@gmail.com}
\email{Ranga-Ram Chary: rchary@caltech.edu}
\author[0000-0002-6039-8212]{fan yang}
\affil{National Astronomical Observatories, Chinese Academy of Sciences,\\ 20A
Datun Road, Chaoyang District, Beijing 100101, China \\}
\affil{IPAC, Caltech, KS 314-6, Pasadena, CA 91125, USA}
\affil{School of Astronomy and Space Science, University of Chinese Academy of Sciences,
Beijing 100049, China\\}
\affil{Department of Astronomy, Beijing Normal University, Beijing 100875, People's Republic of China\\}

\author[0000-0001-7583-0621]{ranga-ram chary}
\affil{IPAC, Caltech, KS 314-6, Pasadena, CA 91125, USA}
\author[0000-0002-2874-2706]{ji-feng liu}
\affil{National Astronomical Observatories, Chinese Academy of Sciences,\\ 20A
Datun Road, Chaoyang District, Beijing 100101, China \\}
\affil{School of Astronomy and Space Science, University of Chinese Academy of Sciences,
Beijing 100049, China\\}
\date{April 2019}
\begin{abstract}
We present a re-analysis of transit depths of KELT-19Ab, WASP-156b, and WASP-121b, including data from the Transiting Exoplanet Survey Satellite (TESS). The large $\sim$21$\arcsec$ TESS pixels and point spread function result in significant contamination of the
stellar flux by nearby objects. We use Gaia data to fit for
and remove this contribution, providing general-purpose software for this correction. We find all three sources have a larger inclination, compared to earlier work. For WASP-121b, we find significantly smaller values (13.5 degrees) of the inclination when using the 30 minutes cadence data compared to the 2 minutes cadence data. 
Using simulations, we demonstrate that the radius ratio of exoplanet to star ($R_{p}/R_{\ast}$) is biased small relative to data taken with a larger sampling interval although oversampling corrections mitigate the bias.
This is particularly important for deriving sub-percent transit differences between bands. We find 
the radius ratio of exoplanet to star ($R_{p}/R_{\ast}$)
in the TESS band is 7.5$\sigma$ smaller than previous work for KELT-19Ab, but consistent to within $\sim$2$\sigma$ for WASP-156b and WASP-121b. The difference could be due to specific choices in the analysis, not necessarily due to the presence of atmospheric features. The result for KELT-19Ab possibly favors a haze-dominated atmosphere. We do not find evidence for the $\sim$0.95\,$\mu$m water feature contaminating transit
depths in the TESS band for these stars but show that with photometric precision of 500ppm and with a sampling of about 200 observations across the entire transit, this feature could be detectable in a more narrow $z-$band.

\end{abstract}


\section{Introduction}
\label{sect:intro}
Exoplanet transit depths in multiple bands in the light curve of a star provide clues into understanding the atmospheric composition of the exoplanet \citep{Vidal2003,sing2011,Berta2012}.  
The transmission spectrum leads to the detection of certain atomic and molecular species, as well as haze and clouds in the planet atmosphere
\citep{seager2010}. For instance, the transit spectrum of HD 209458b and XO-1b provides evidence for water and sodium absorption \citep{Deming2013}, while
\citet{sing2016} found evidence for water vapor in hot Jupiters. Similarly, WASP-12b has been characterized as a prototypical hot Jupiter with a high C/O ratio and significant presence of aerosols in its atmosphere \citep{Madhusudhan2011, sing2011}. However, many of these analyses rely on {\it Hubble} Space Telescope spectroscopy. {\it Hubble}
 due to its 90 min orbit, and passage through the South Atlantic Anomaly does not provide a good sampling of the light curve during ingress and egress. Those data are crucial for constraining the inclination of the orbit and thereby the relative transit depths in different bands.

The recently commissioned Transiting Exoplanet Survey Satellite \citep[TESS]{Ricker2015} offers high precision photometric measurements in a broad optical band ($0.6-1$\,$\mu$m) that is redward of the band used by {\it Kepler}. TESS aims to discover transiting exoplanets around the brightest stars in the vicinity of the Sun. It has four cameras with a total field of view 24$\times$96 degrees. The 50$\%$ ensquared-energy (within a square) half-width is 1 detector pixel or 21 arcseconds. The 90$\%$
ensquared-energy half-width is 2x2 pixels. TESS produces multi-frame images at a cadence of 30 minutes with a baseline of at least 27 days
and thus does not suffer from gaps in the light curve like {\it Hubble}. The photometric precision is about 1$\%$ sensitivity at 16th mag. 

Here we present a re-analysis of the exoplanets
KELT-19Ab, WASP-156b, and WASP-121b
by combining TESS with other past observations. This paper is organized as follows. Section 2 describes the generation of TESS light curves, derivation of planet parameters from the TESS data using a Monte-Carlo Markov Chain (MCMC) method, transit parameters bias caused by binning and a comparison between our values of inclination and $R_{p}/R_{\ast}$ with previous estimates. Constraints on the exoplanet atmosphere and the predictions of $R_{p}/R_{\ast}$ in the Kepler band and a $z-$band, which is sensitive to water vapor is presented in Section 3.
In Section 4, we present a summary of our results.

\section{Data reduction}
\subsection{TESS Photometry }
The exposure time of TESS for each frame is 2 seconds. Due to
data downlink limitations, two images products are released; the individual frames in a few pixels around certain sources are integrated to 2 minute cadence, and are called Target Pixel Files (TPF). The frames are also integrated to 30 minute cadence for all the sources in the field of view, to generate a Full Frame Image (FFI) \citep{2015Ricker}. We use 2 minute cadence frames to derive the light curves and 30 minute cadence frames to calibrate and remove the blending flux from surrounding sources.

The FFI shows an offset in astrometry relative to the nominal position of the exoplanet host star, as measured by Gaia. However, the astrometric offset is similar across all the frames. So for each source, we checked the astrometry in a few frames and applied a-few-pixel-level offset if it is present. A smaller fractional offset is
determined from a 2-dimensional Gaussian fit of the target. Example images after correcting for astrometric offsets are as shown in Figure 1.
The numbers of frames for KELT-19Ab, WASP-156b, WASP-121b are 17612, 17612, and 15973, respectively. The median offsets for KELT-19Ab, WASP-156b, WASP-121b are 0.10, 0.27, 0.07 in pixels. The standard deviations are 0.04, 0.09, 0.04 in pixels. We apply the same astrometry correction when we analyze the 2 minute cadence image.

The Target Pixel File (TPF) is a 11 pixels $\times$ 11 pixels
cutout of the image around the TESS Input Catalog (TIC). The pipeline versions are spoc-3.3.57-20190215 for KELT-19Ab, spoc-3.3.51-20181221 for WASP-156b, spoc-3.3.57-20190215 for WASP-121b.

\begin{figure}[ht]
  \centering
  \includegraphics[width=3.5in]{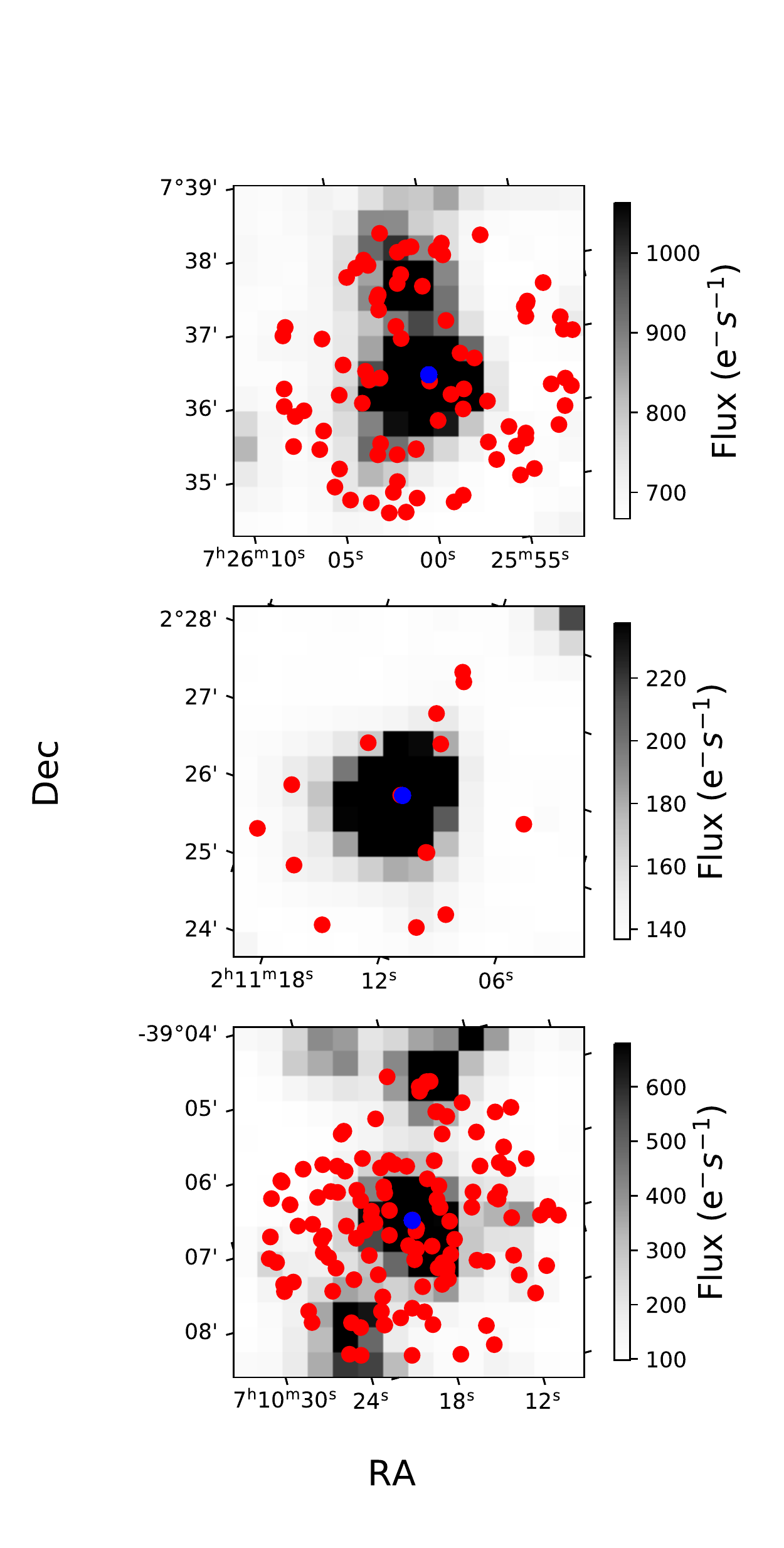}
  \caption{14$\times$14 pixel TESS images of KELT-19Ab, WASP-156b, WASP-121b, from top to bottom, respectively. The blue markers indicate the position of the exoplanet host star in the Gaia catalog. The red markers indicate the position of nearby sources in the Gaia catalog which contaminate the photometry of the host star i.e. within 6 pixels separation. Each pixel in the image is 21$\arcsec$. The direction is North up, East to the left.  
  }
\end{figure}

We measured the photometry of the host star in a circular aperture with TPF. The aperture radius was chosen to be 3 pixels, corresponding to 63$\arcsec$. The flux from pixels whose fractional area falls within the aperture is correctly accounted for.
The sky background per image was estimated as the median of the pixels which constitute the lowest fifth percentile in flux and subtracted
from the photometry in the aperture. The standard deviation of these pixels yields the background noise in the photometry. The quadrature sum of this noise and the Poisson noise of the source itself is the photometric uncertainty of the data point.
The contribution to the photometry from nearby unresolved stars was then removed based on the relationship between flux brightness profile and the distance to the Gaia centroid of the unresolved stars (details in the next section).

\subsection{Correction for Blending Sources}  

We then corrected the photometry to remove the contamination from surrounding unresolved sources. We used external information from the Gaia database \citep{gaia2016,Gaia2018}, including position, brightness in Gaia G and Rp bands. The Gaia Rp band is similar to the bandpass of TESS. However, it is less sensitive than the Gaia G band. For stars which have Rp band photometry from Gaia, we used its brightness information directly. 
For faint stars with no Gaia Rp band detection, we used their G band flux density and applied a median color correction which is the median flux density ratio between the Rp band and G band of stars in the vicinity.

In order to estimate the flux from these contaminating sources inside the aperture centered on the exoplanet host star, we used the TESS FFI data to derive the relationship between flux contamination and the distance between the center of the photometric aperture and a source. We chose 5 bright (brighter than 13 mag in Gaia G-band) and isolated sources from the TESS Input Catalog. It is difficult to guarantee complete isolation to within 6 TESS pixels but the bright stars were chosen to have
no detectable companions in the image and the Gaia catalog showed an absence of sources within 5 mags of the bright stars. We divided the 6 pixel radius circle into a set of annuli, each of 0.25 pixel width. We then chose a set of positions that traverse
the annulus circumference in 0.1 pixel steps. With each of these positions as center, we performed aperture photometry by
measuring the flux density within a circular aperture of radius 3 pixels as for the exoplanet host star. Sky subtraction was performed as described in the previous section.
The maximum separation between the center of the aperture and from the bright star position was allowed to be 6 pixels where the contamination is negligible. This is because
TESS has a 90\% ensquared energy within a half-width of 2 pixels.

Finally, we normalized the set of sky subtracted flux density measurements by the flux measured at zero separation. We repeated this for 5 targets and for each target we used 10 frames. The median and standard deviation of the normalized brightness profile for all these measurements is shown in Figure 2.

The relationship between distance and brightness profile was then estimated by the polynomial fitting of the data pairs (see in Figure 2). We then apply
this relation to every contaminating source whose brightness has been estimated as described above, to derive the flux density contributed by the blending source to the photometry of the exoplanet host star. This contaminating flux density, which is constant for a particular exoplanet host star, was then removed from our photometry values\footnote{
The code to apply this correction is available at \url{https://github.com/sailoryf/TESS_Deblending/
}}. The total contaminating flux percentage and uncertainty for the three stars is
 6.22$\%$$\pm$0.42$\%$, 0.64$\%$$\pm$0.03$\%$ and 31.49$\%$$\pm$1.79$\%$ for KELT-19Ab, WASP-156b and WASP-121b respectively.

In addition, there is a possibility of blending from an unresolved binary companion. 
In the case of WASP-156b and WASP-121b, there is no evidence of a binary companion
in past work \citep{Delrez,Demangeon}. In the case of KELT-19Ab, a nearby stellar companion was detected with the Palomar/Hale 200$''$ telescope using the near-infrared adaptive optics (AO) system \citep{KELT-19}. The stars have a measured magnitude difference of $\Delta J=2.50\pm0.06$ and $\Delta K_{s}=2.045\pm0.03$. 
\citet{KELT-19} report a flux ratio of 0.0270$\pm$0.0034 at a wavelength of 5200$\pm$150$\AA$, effective temperatures of 7500$\pm$200K for the primary component, and 5200$\pm$100K for the secondary component. We search the ATLAS library \citep{Kurucz1993} grid based on the flux ratio and effective temperatures. The best fit stellar templates
are multiplied by the TESS filter response to estimate the brightness of the companion star and its contamination to 
the TESS photometry. The derived contaminating flux fraction is 0.059$\pm$0.006 which we remove.

\begin{figure}[h]
  \centering
  \includegraphics[width=3.8in]{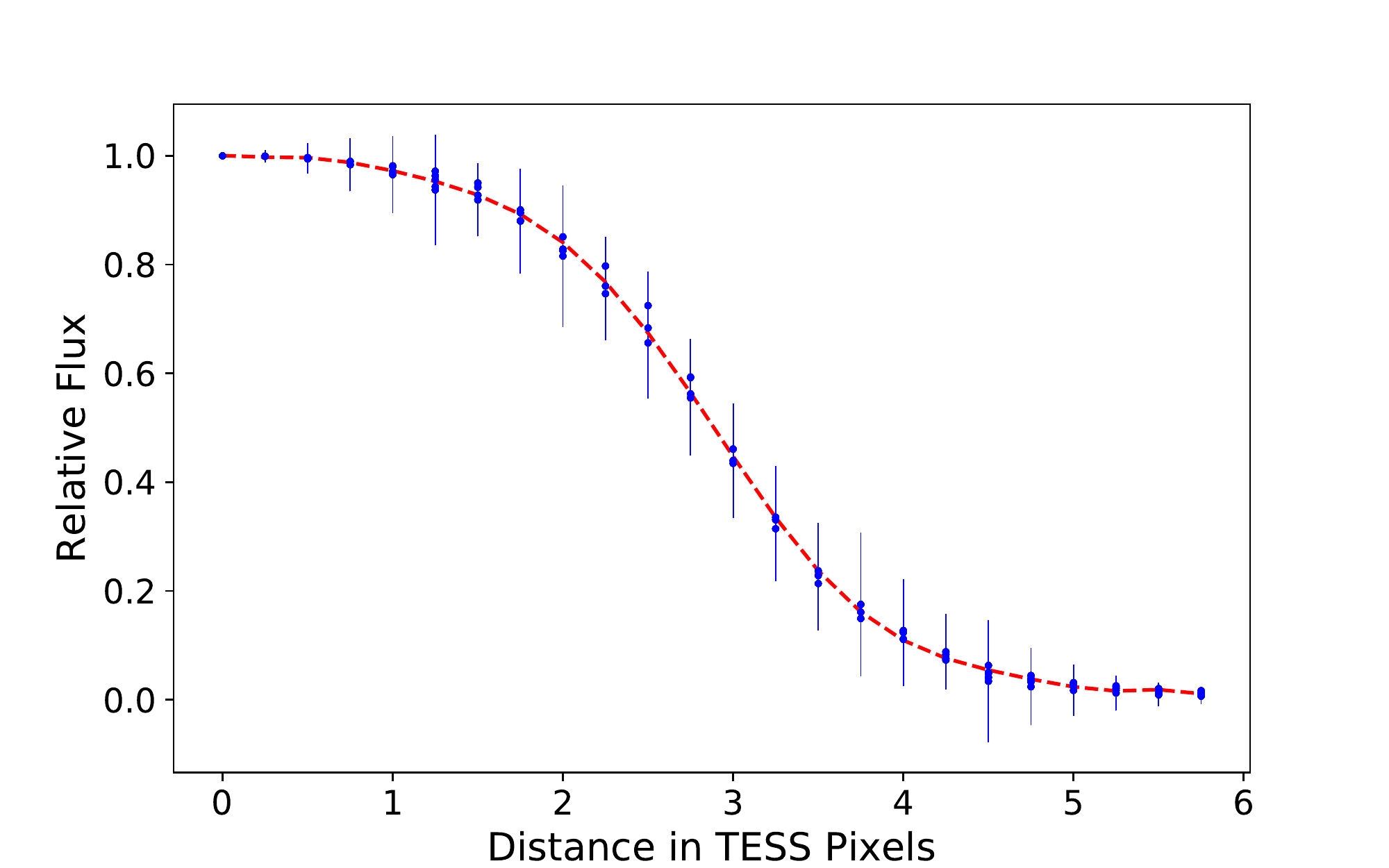}
  \caption{The normalized brightness profile of stars as seen by TESS used to estimate the amount of contamination of an exoplanet host stars' photometry by background stars seen by Gaia. The scatter points are the observed data. The red line indicates the polynomial fit to the data.}
  
\end{figure}

\subsection{Detrending of the Light Curve And Comparison with TESS-PDC Light Curve}

Measuring an accurate transit depth of an exoplanet requires a high precision light curve.
This requires detrending to remove the long term trends in the photometry outside of transit events \citep{GPdetrending,AGP,SFF,PDC}. Since we already know the existence of an exoplanet around 
these stars, we can use prior knowledge of the transit mid-point time, period and transit duration to perform
a simple detrending.
To do this, we predicted the transit time in the TESS light curve and extract the parts of the light curve spanning 0.6 days centered at the
transit mid-point.

The data points of the light curve during the planetary transit were first masked out. We fitted a linear function to the signal in the masked, cut-off, light curve. The fit is used to estimate and remove the trending in the light curve
within a day. We applied this correction for every single data point of the extracted light curve and fold the light curve around the transit midpoint to
get the transit events shown in Figure 3. We also tried higher order polynomial functions, including orders of 2, 3, 5 and 10. The difference in the detrended light curve is negligible for these choices of polynomials.

The photometric data reduction could be influenced by factors already corrected in our pipeline, or by some factors beyond the consideration of our work, e.g., point spread function (PSF) variation, focus changes, bad pixels, and thermal variations. In order to make sure these factors do not dominate the precision of transit depth measurements, we compare
our results with light curves derived using the TESS Pre-Search Data Conditioning (PDC) modules. PDC light curve is produced by TESS collaboration with consideration of factors above. The comparison covers three targets in this work, as well as HD-219666, an isolated exoplanet host star identified by TESS which is bright compared to its surrounding stars \citep{Esposito2019}. 
The deblending correction for HD-219666 is 0.212$\%\pm0.006\%$.

 We compare the median flux of our light curve with the G-band flux reported by Gaia \citep{Gaia2018}. The flux density ratio of our decontaminated flux to Gaia flux for KELT-19Ab, WASP-156b, WASP-121b and HD-219666 is 97, 71, 90, and 80, respectively (in arbitrary units). The standard deviation is 9.7. The flux ratio of PDC flux to Gaia flux is 95, 66, 86, 75, respectively. The standard deviation is 11.0. This indicates that the two techniques are perfectly consistent with each other. Furthermore, we find that
the derived planetary transit depths from the two sets of light curves of these sources are within 1 $\sigma$ of each other.

\subsection{Planet Parameters Derived from TESS-Detected Transits}
\subsubsection{Fitting with Free Inclination and Semi-major Axis}
We used an MCMC fitting technique to derive the parameters of the exoplanets from the TESS transit curves \citep{Mandel_Agol2002}. We adopted a
circular orbit as in previous articles that confirmed the corresponding exoplanets with orbital periods derived therein \citep{KELT-19,Demangeon,Delrez}. 
The value of T$_{eff}$, log $g$, and [Fe/H] was taken from the
discovery articles as well.
The free parameters in our fit are the radius ratio of the planet to the host star ($R_{p}/R_{\ast}$), the inclination of the planets orbit (i), the semi-major axis in unit of stellar radii ( $a/R_{\ast}$), time offset of transit center (T0) and the limb darkening parameters (a1 for linear limb-darkening coefficient, a2 for quadratic limb-darkening coefficient), as shown in table 1. Except for limb darkening, the other parameters have uniform priors.

We adopted the common inclination prior of $\mathcal{U}$[70,90] but changed it to be $\mathcal{U}$[70,110] when the posterior of inclination is within a degree of 90 degrees. This is because the distribution of inclination performs differently when the real inclination is close to 90. A prior cut-off at 90 degree severely distorts the posterior distribution to be non-Gaussian. A prior with no 90-degree cut-off yields a more reasonable posterior distribution. The difference in inclination posterior between these two scenarios of priors is systematic, but consistent within 1 $\sigma$.

The limb darkening model is the same as that adopted in the exoplanet discovery articles \citep{KELT-19,Demangeon,Evans}. 
The coefficients for limb darkening have a Gaussian prior centered on the prediction for TESS data \citep{TESSLD} with an uncertainty of 0.05 \citep{YangLD}.

\begin{figure}[!t]
  \centering
  \includegraphics[width=3.5in]{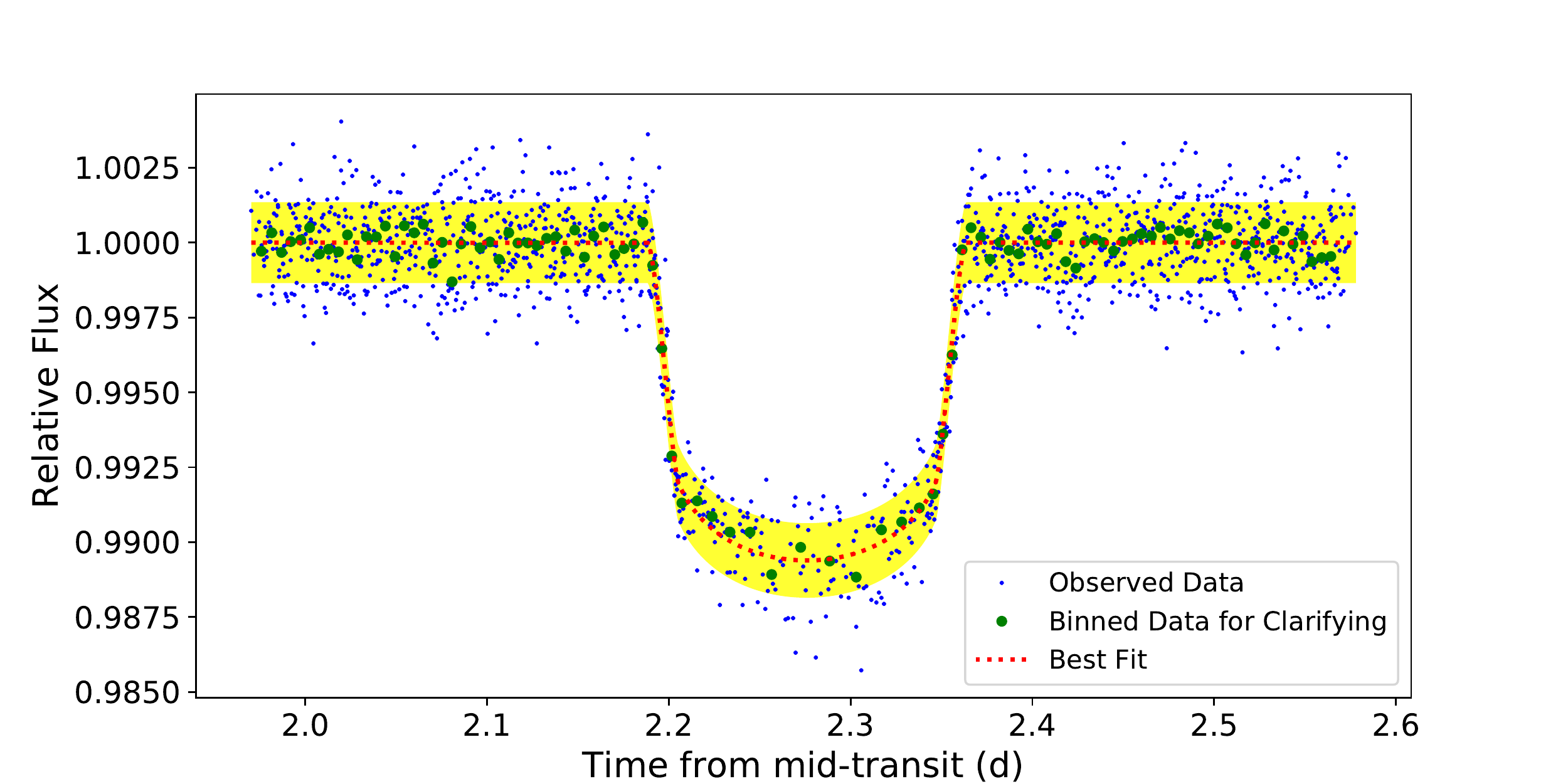}
  \includegraphics[width=3.5in]{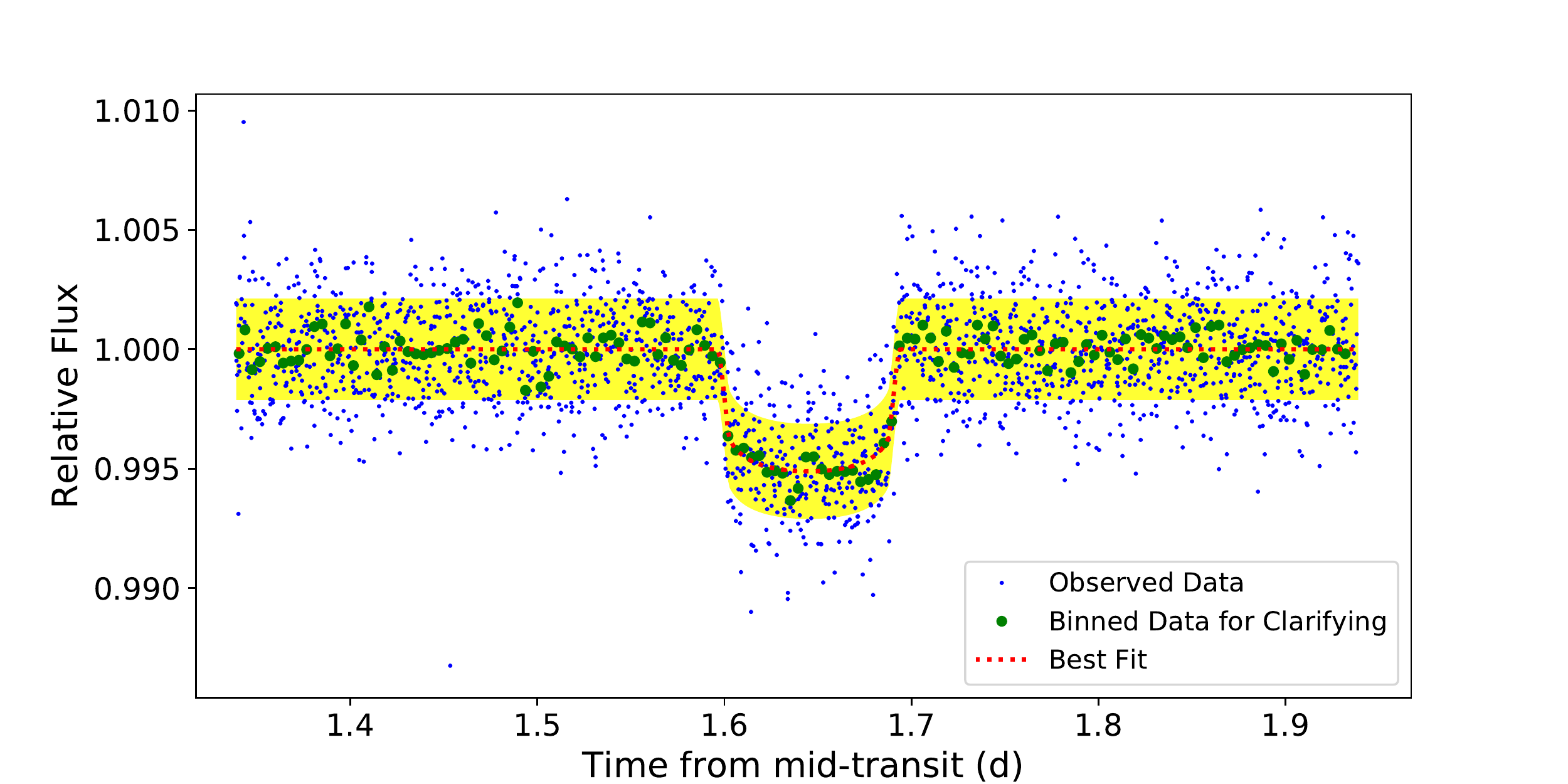}
  \includegraphics[width=3.5in]{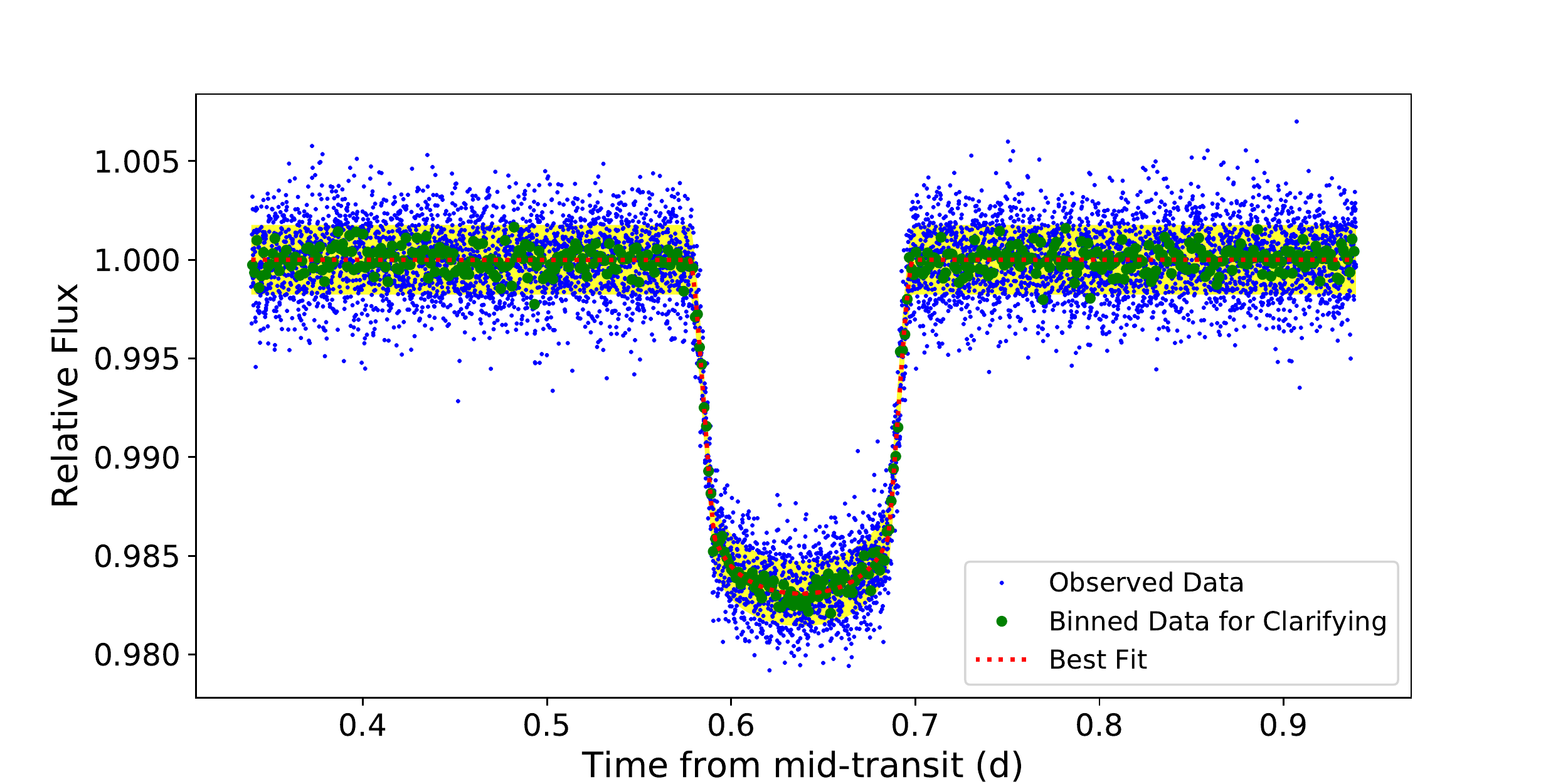}
 \caption{Fits to the TESS TPF transit light curves of KELT-19Ab (top), WASP-156b (middle), WASP-121b (bottom). The blue points are observed 2 minute data decontaminated using Gaia photometry. The green points are bins of 30 minutes cadence data. The red dashed line represents the best-fit transit model to the 2 minute data. The yellow region shows 1 $\sigma$ confidence region of the fits. The midpoint of the transit event is shifted to half the period, at the center of the x-axis. The free parameters in the fits are inclination, semi-major axis and limb darkening parameters as shown in Figure 4.}
\end{figure}

In the case of KELT-19Ab, we adopted T$_{eff}$, log \emph{g}, and [Fe/H] of 7500$\pm$110 K, 4.127$\pm$0.029 cgs, -0.12$\pm0.51$. As in \citet{KELT-19}, a quadratic limb darkening model is applied. The priors for the limb darkening coefficient are 0.27$\pm$0.05, 0.23$\pm$0.05, for the linear (linLimb) and quadratic coefficients (quadLimb), respectively.

In the case of WASP-156b, the T$_{eff}$, log \emph{g}, and [Fe/H] values are 5871$\pm$57 K, 4.35$\pm$0.03 cgs, 0.10$\pm0.10$. A quadratic limb darkening model is used \citep{Demangeon}.
The priors for limb darkening coefficients are 0.36$\pm$0.05 and 0.22$\pm$0.05 for linLimb and quadLimb respectively.

\begin{figure}[!htb]
  \centering
   \includegraphics[width=3in]{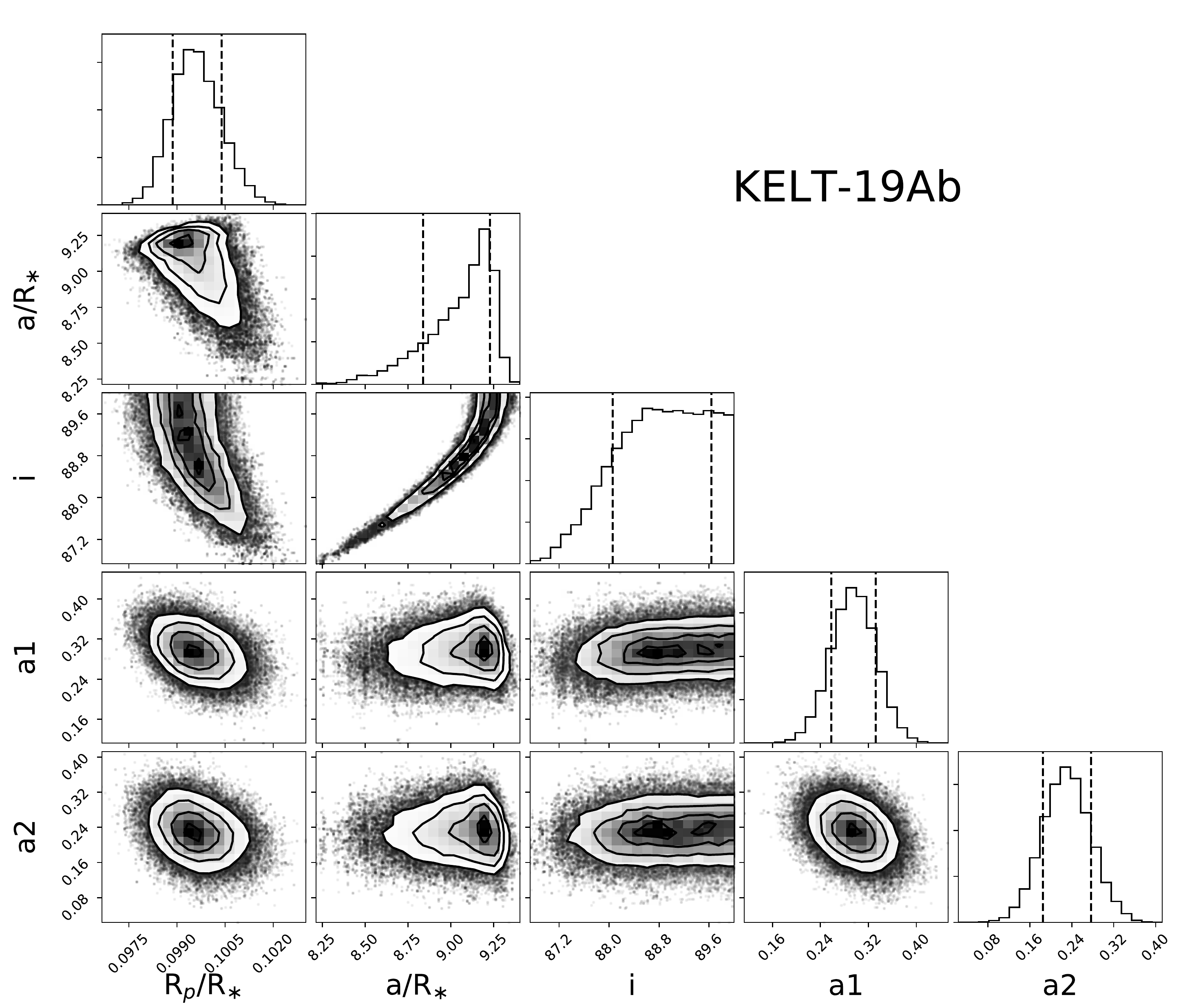}
   \includegraphics[width=3in]{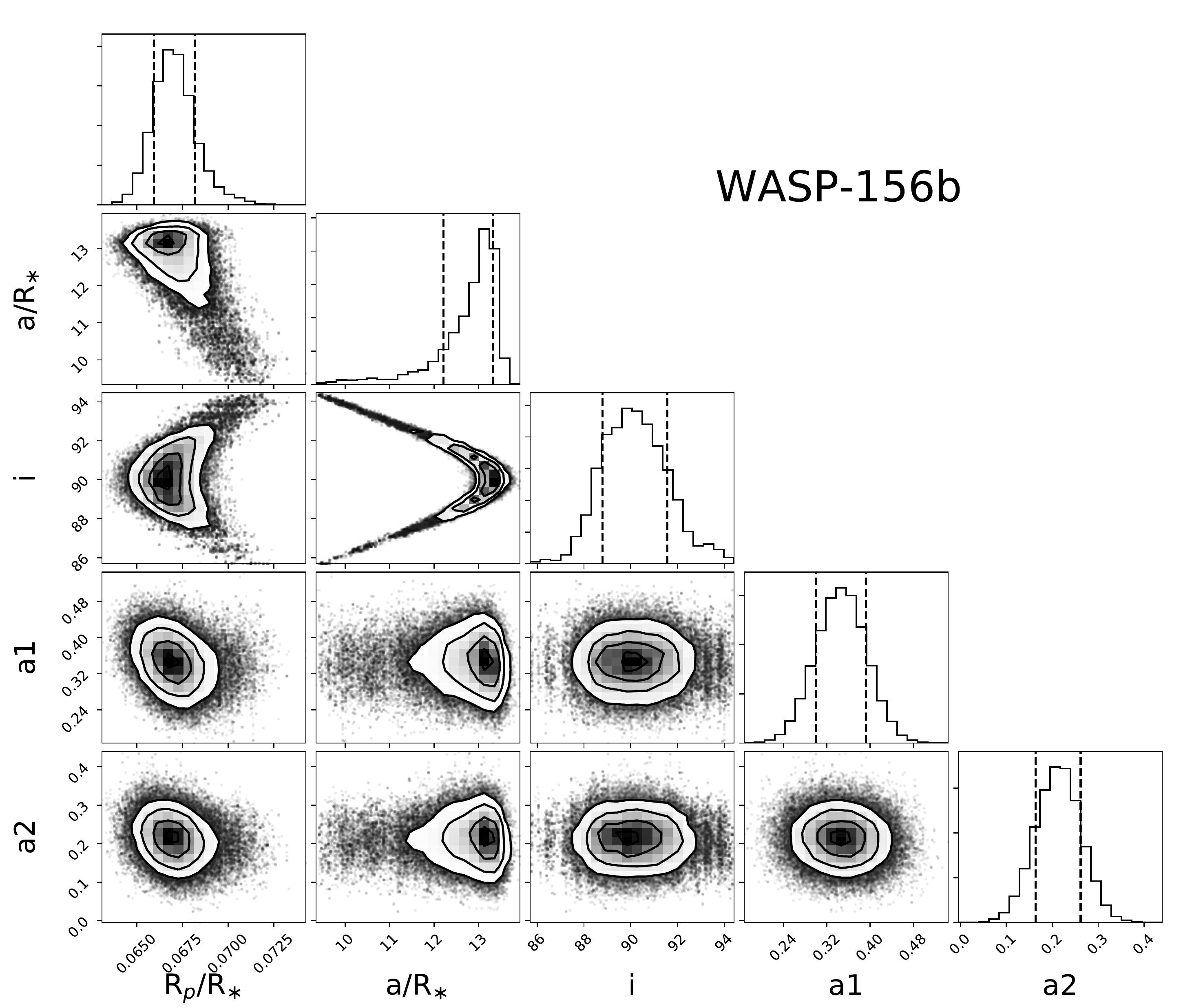}
   \includegraphics[width=3in]{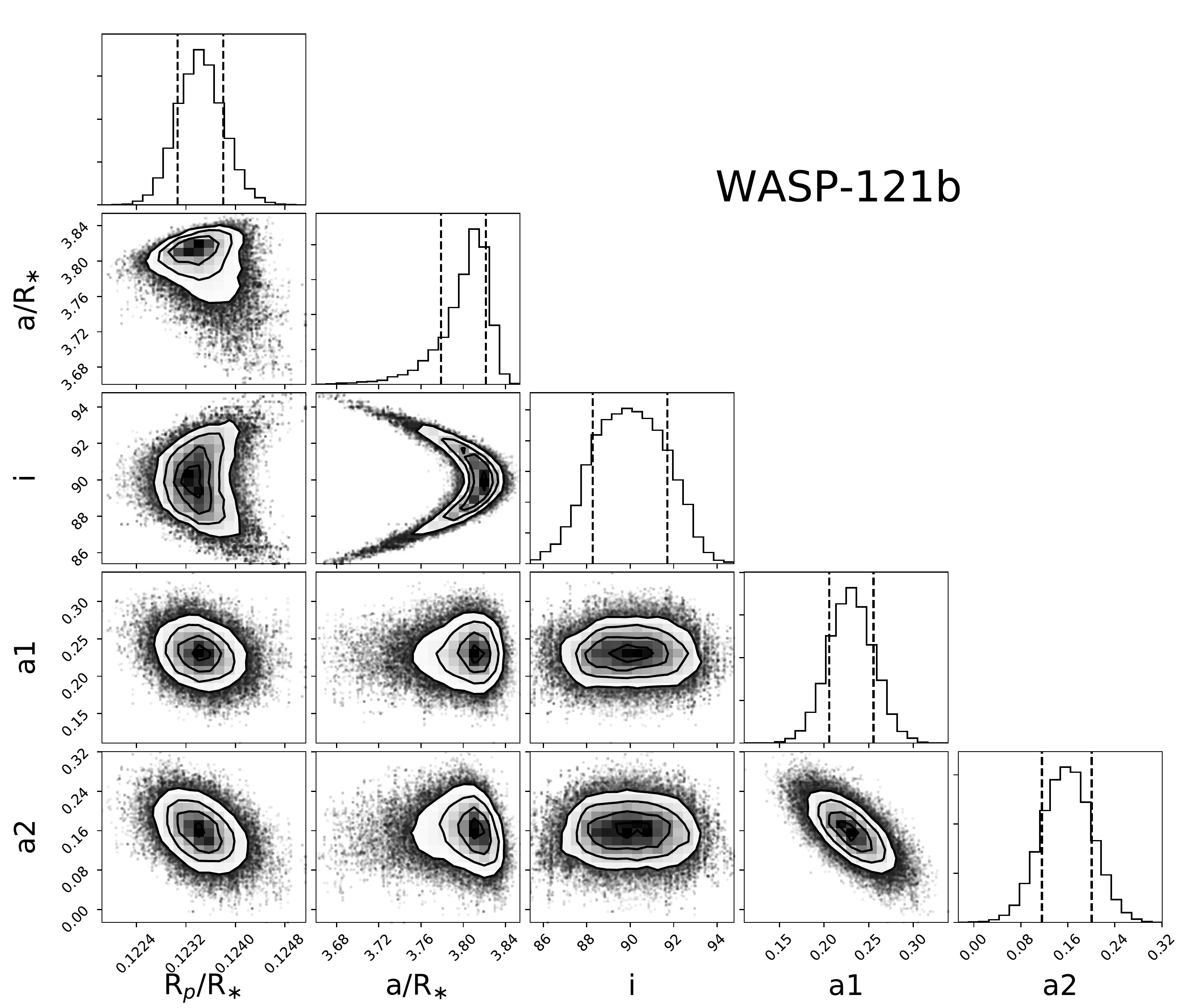}
  \caption{The probability distribution of the fit parameters for KELT-19Ab (top), WASP-156b (middle), WASP-121b (bottom). The grayscale map indicate the density of samples derived from MCMC analysis for different pairs of parameters. The histograms along the diagonal are the marginalized density distribution for individual parameters. The dashed lines denote 68$\%$ of samples. The plot applies corner routine \citep{corner2016}. }
\end{figure}

In the case of WASP-121b, the T$_{eff}$, log \emph{g}, and [Fe/H] values are 6460$\pm$140 K, 4.2$\pm$0.2 cgs, 0.13$\pm0.09$. The limb darkening model used is the quadratic limb darkening model \citep{Delrez,Evans2018}. The limb darkening coefficients that we applied to our MCMC fits are 0.33$\pm$0.05, 0.21$\pm$0.05, respectively.

For each fitting, the parameters' probability distributions are derived using Python MCMC tool, PYMC \citep{pymc}. We applied 200000 iterations for the whole chain, and ignore the first 100000 steps to ensure the chain is stable. The best-fitting light curves are shown in Figure 3, respectively. The distribution of $R_{p}/R_{\ast}$, $a/R_{\ast}$, inclination and limb darkening parameters are shown in Figure 4.
The best fit results are shown in Table 1.

The inclinations derived from TESS TPF data for KELT-19Ab, WASP-156b, and WASP-121b are 3.3, 1.0 and 2.3 degrees larger than the results based on identification paper, respectively. Specifically, in case of WASP-121b, the inclination derived is more consistent with the follow-up work value 89.1 by \citet{Evans2018}.
Due to TESS's continuous sampling with 2 minutes and high precision photometry, the ingress and egress of transit are well monitored. These short phases in the transit duration are extremely sensitive to the inclination and semi-major axis. We, thereforebelieve that the TESS result as well as result from \citet{Evans2018} on inclination for WASP-121b are robust.

\subsubsection{Fitting with Fixed Inclination and Semi-major Axis}

Although our best fits are similar to other work, there are clear signs of differences, particularly in our derived inclination and transit depths.
In order to focus on differences in the radius ratio caused by wavelength dependent properties, we refit the TESS light curves, by fixing the inclination and semi-major axis to the referenced work \citep{KELT-19,Demangeon,Evans}.

\begin{figure}[!htb]
  \centering
   \includegraphics[width=3.5in]{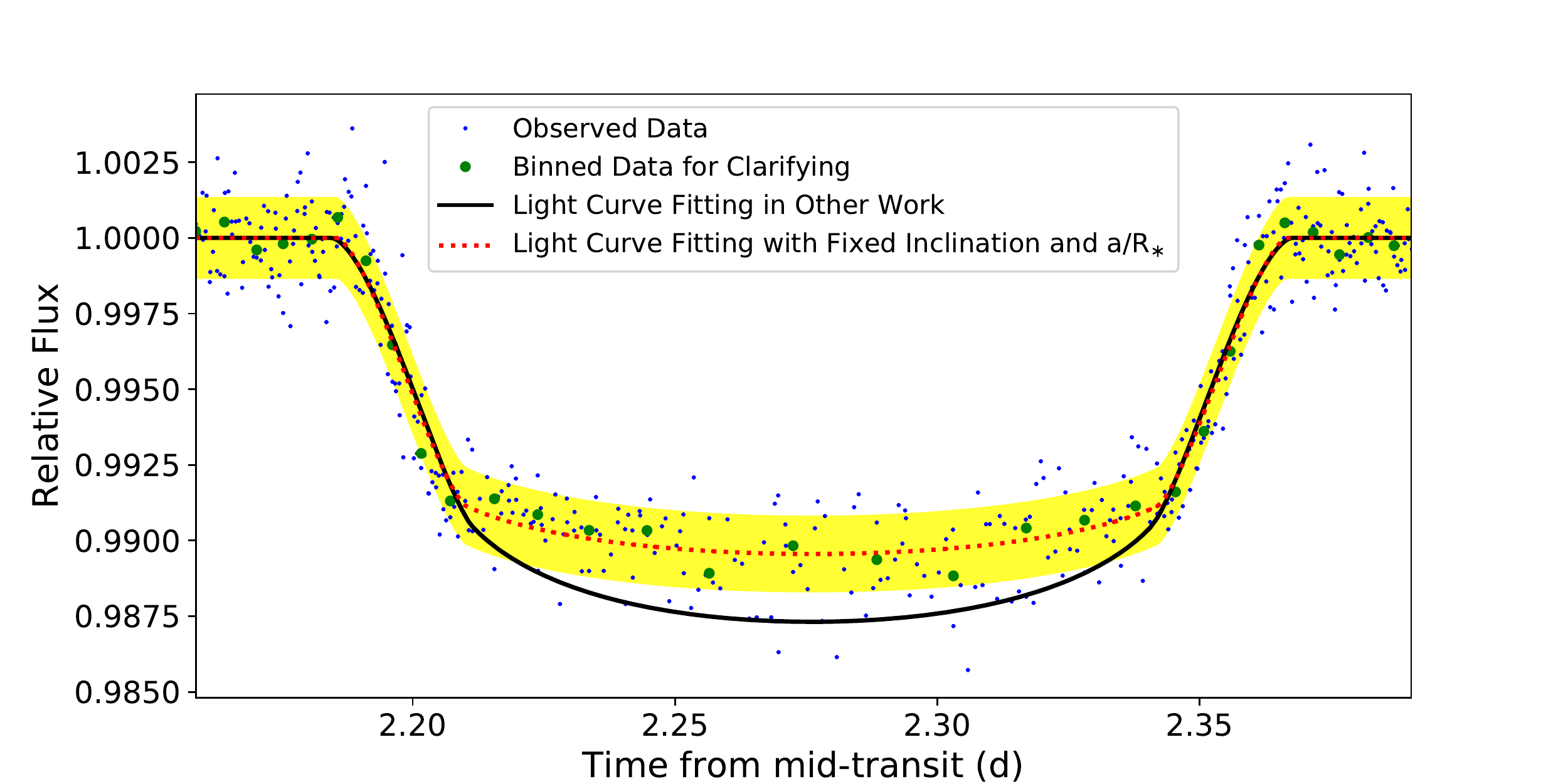}
   \includegraphics[width=3.5in]{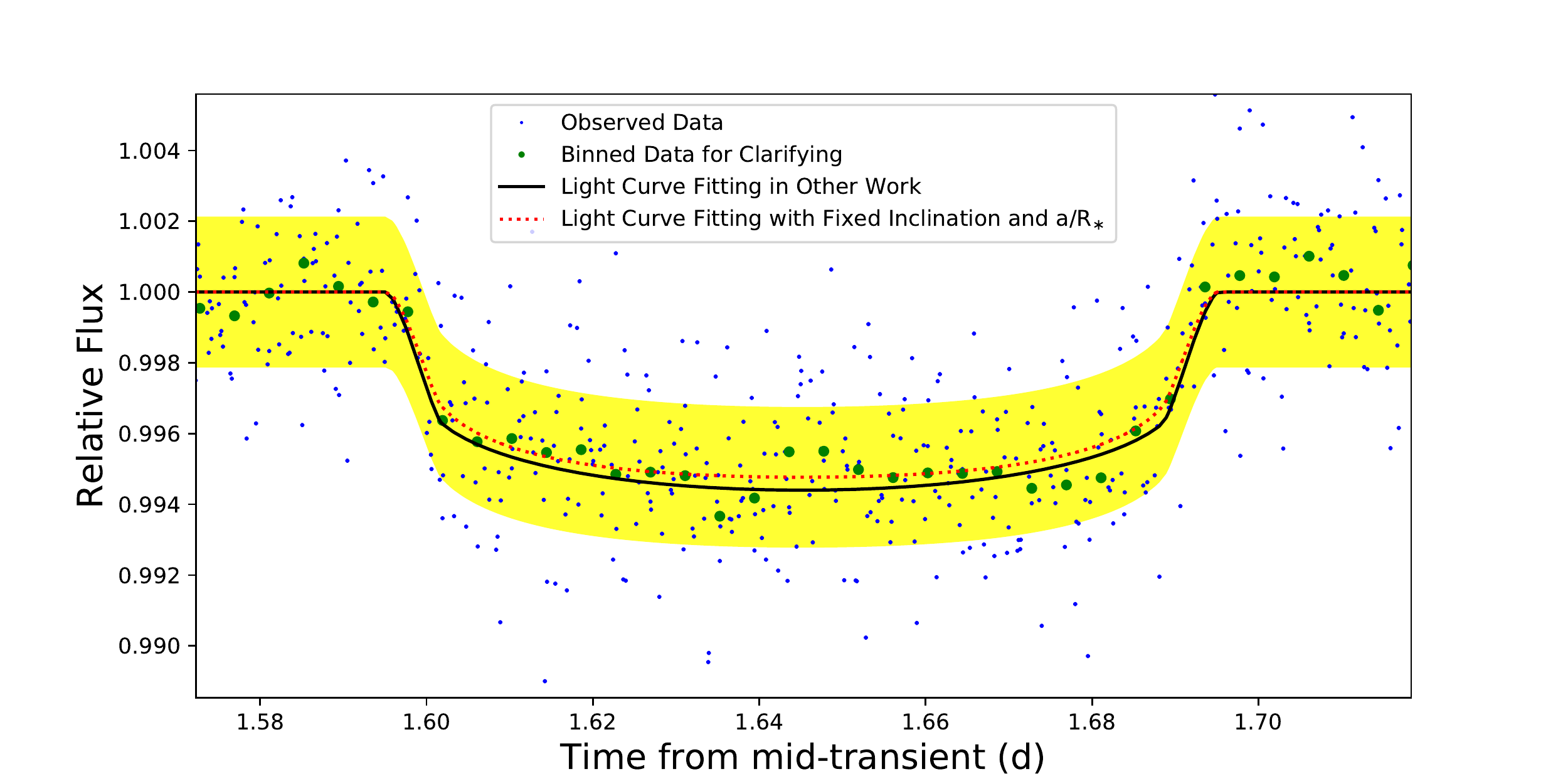}
   \includegraphics[width=3.5in]{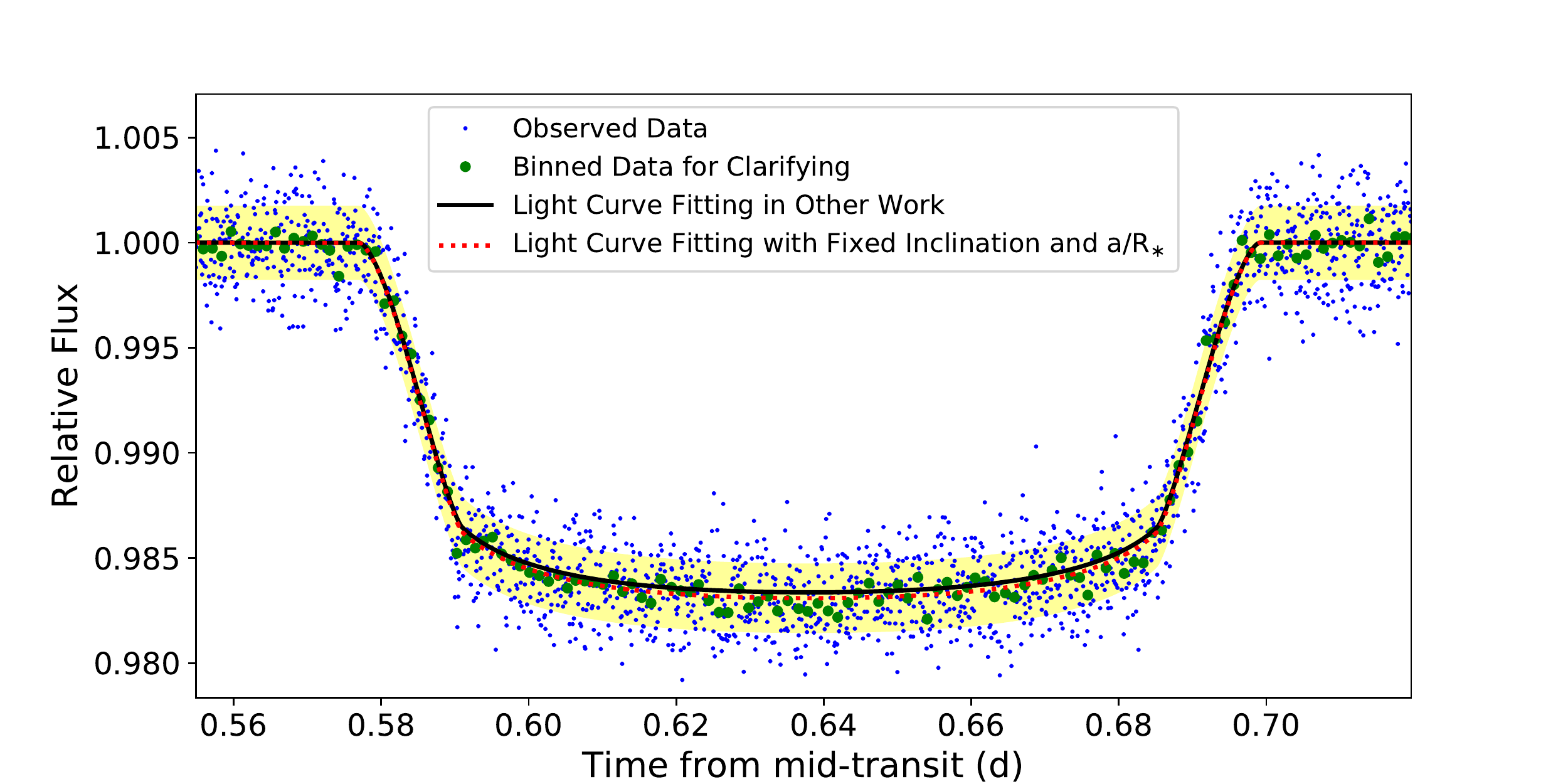}
  \caption{Results from fitting with fixed inclination and semi-major axis for KELT-19Ab (top), WASP-156b (middle), WASP-121b (bottom). The blue points are deconfused photometry. The green points indicate 30 minutes cadence binned result. The red dashed line represents best-fitting model. The black line indicates the result of previous work (referenced in the text). The yellow region shows 1 $\sigma$ confidence region of the fitting. The transit event is shifted at the center of x-axis which is half of the period.}
  
\end{figure}

The fitting followed the same procedure as section 2.4.1. The fitting results, as well as the derived parameters from previous work, are listed in Table 1. The differences between our fits and the result from the former work
are shown in Figure 5. The reduced chi-square in the fits are slightly larger than the best fit from previous section (Table 1). The small difference in the chi-square values indicate that the value of inclination and semi-major axis preferred in other work, is reasonably consistent with the TESS data but is not the preferred solution. 

The fixed parameter MCMC fits have reduced uncertainties due to the strong priors on inclination and semi-major axis. These constraints on inclination and semi-major disturb the Gaussian distribution of other parameters. The revision for the three sigma rules for the parameters is beyond this work \citep[see in reference][]{Friedrich1994,MacKay2002,Hogg2018}. We cited the 1$\sigma$ uncertainty from the MCMC derived standard deviation for each parameter.

Our fits to the TESS data gives us a consistent planet to star radius ratio for WASP-121b and WASP-156b at the $\sim$2$\sigma$ level (Table 1). However, the planet to star radius ratio for KELT-19Ab is 7.5$\sigma$ smaller than previous work \citep{KELT-19}. The TESS radius ratio for WASP-121b is deeper than the median value including all the bands with the TESS wavelength range from \citep{Evans2018}. For WASP-156b, the radius ratio is smaller than past work \citep{Demangeon}.

\subsection{ Transit Parameters Bias Caused by Binning }
Sampling cadence results in morphological distortions to the transit light curve due to sampling of transit ingress and egress \citep{David2010}. For the TESS cadence, an assessment of the parameter bias is necessary for assessing the differences in transit depths for different bands.

We firstly investigate if our derived parameters from
the TESS 2 minute data are consistent with that from the 30 minute data by repeating the whole data reduction and fitting with 30 minute cadence image. In case of free inclination fitting, the inclinations derived for KELT-19Ab, WASP-156b, and WASP-121b are 4.6, 5.1 and 13.6 degrees smaller than the results based on 2 minutes binning data, respectively. Furthermore, the residuals from the light curve for WASP-121b when fixing the inclination shows obvious structure during the phase of ingress and egress (seen in Figure 6). The radius ratio when fixed inclination and semi-major axis is 0.1195$\pm$0.0004, about 7$\sigma$ smaller than 2 minutes result, with substantially worse chi-square values. 
The inconsistency, especially for WASP-121b, motivates a simulation for a more detailed assessment of the differences.

\begin{figure}[!htb]
  \centering
    \includegraphics[width=3.5in]{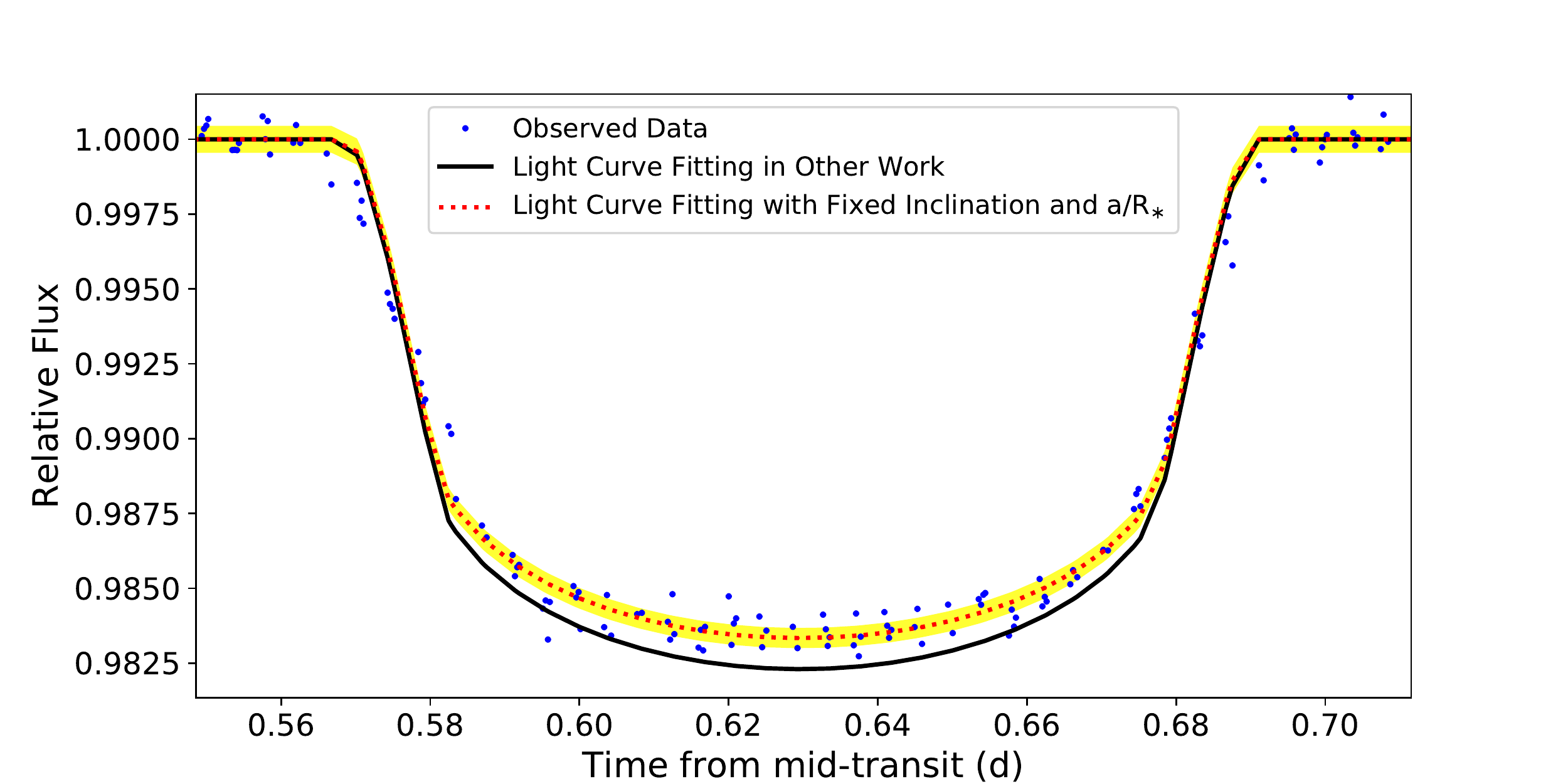}
    \includegraphics[width=3.5in]{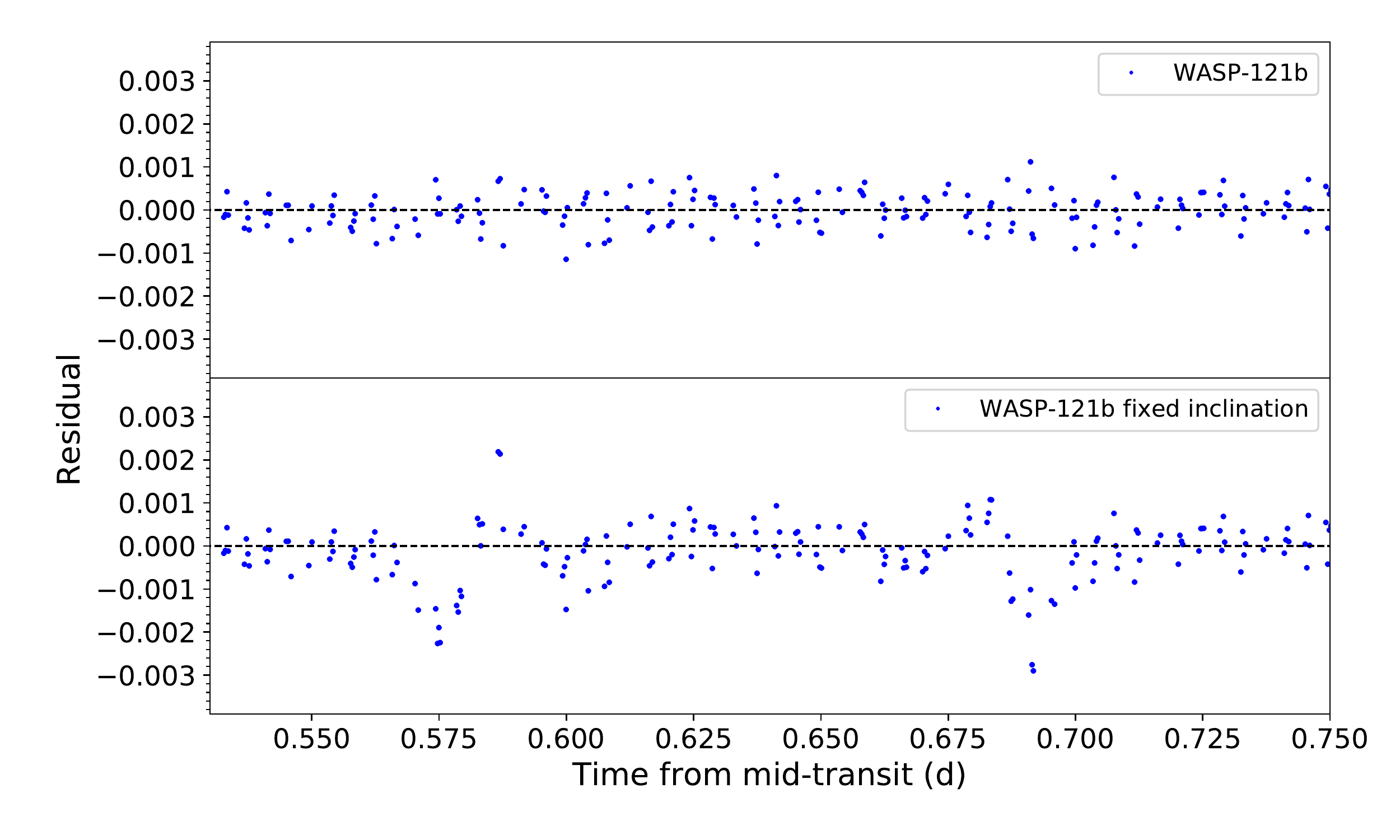}
  \caption{Top: Result from fitting with a fixed inclination and semi-major axis for WASP-121b (TESS 30 minute data). The red dashed line represents the best-fitting model. The black line indicates the fit from previous work. Bottom: Residuals after fitting for the exoplanet transit around transit midpoint for WASP-121b based on 30 minutes data. The upper panel is residuals from the best fit. The lower panel is residuals assuming a fixed inclination and semi-major axis adopted
  from previous work. The 30 minutes data are clearly
 inconsistent with the previously derived inclination value.}
\end{figure}

We simulated 1000 light curves for each source. The input parameters were set to that in the discovery papers \citep[Table 1;][]{KELT-19, Demangeon, Delrez}. We note that for WASP-121b, the newest reported inclination is 1.5 degrees larger than initially assessed \citep{Evans2018}. The result from \citet{Evans2018} is also well matched  to the TESS TPF derived result as shown in Table 1. But for the purpose of the simulations, we chose the value from the discovery paper. The time sampling for the simulated light curve is 1 second and is binned to 2 minutes and 30 minutes. The numbers of transit events are set to be the same as the event numbers observed by TESS. We add a Gaussian error to the light curves. The adopted 1$\sigma$ scatter is the standard deviation of the residuals from fitting the TESS light curves scaled by a factor of the square root of time scale.

These simulated light curves are then fit as described earlier.
Figure 7 and 8 show the distribution of derived parameters compared
to the input values.
We find that the magnitude of bias in the derived parameters especially inclination and semi-major axis, depends on the duration time of the transit, especially ingress and egress. We take WASP-121b as an illustrative example which gives the largest inclination difference. The derived median inclination value with the 30 minute cadence data is 76.4$\deg$ with a standard deviation of 0.7, 11.2 degrees smaller than the input value. The inclination decrease is because the binning smooths the rapid change of the flux decrease during ingress and egress. The planet to star radius ratio when fixing the inclination and semi-major axis is 0.1201$\pm$0.00035, also smaller than the input value at the level of $\sim5\sigma$ (seen in Figure 7).

\citet{David2010} provide a numerical solution to compensate for biases caused by binning, using Kepler’s equations. 
We investigate the derived correction through a simulation. 1000 simulated light curves of 30 minute cadence 
are generated with input parameters the same as described before. Using the batman routine \citep{Kreidberg2015}, we resample and integrate the simulated light curves to 2 min cadence. The derived fit inclination is 86.4$\pm$1.3 deg, indicating that resampling corrects the inclination bias to within 1$\sigma$. However, the derived R$_{p}$/R$_{\ast}$ from the resampled light curve has a median and scatter of 0.1231$\pm$0.00068. The bias in R$_{p}$/R$_{\ast}$ is 0.0011$\pm$0.00068, while the bias without resampling is -0.0019$\pm$0.00035.

Such differences of $\sim 0.8\%$ are negligible for most transit applications but need to be considered carefully for atmospheric characterization. 
We use further simulations to assess if any R$_{p}$/R$_{\ast}$ bias after oversampling still exists. The resampling cadence is changed from 1 minute to 10 minutes with a step of 1 minute. The resampling cadence below 1 minute is also tested from 10 seconds to 50 seconds with a step of 10 seconds. 
The R$_{p}$/R$_{\ast}$ reveals an offset at the level of 0.001 with an uncertainty $\sim$ 0.0006 when the resampling cadence is below 5 minutes. The offset is larger and increases with resampling cadence when the resampling cadence is larger than 5 minutes.

In addition, we apply simulations to check if the bias still exists when the initial sampling rate is 2 minutes. The resampling cadence is set to be 30 seconds, 1 minute, and 2 minutes. The derived R$_{p}$/R$_{\ast}$ does not reveal any evidence of offset suggesting that the batman routine itself does not induce any systematic bias.
We thus conclude that the resampling technique \citep{David2010, Kreidberg2015} significantly reduces the systematic bias caused by undersampling. However, we conclude that the very high precision (R$_{p}$/R$_{\ast}$ $\sim$ 0.5$\%$) needed for atmospheric analysis needs careful correction for sampling-induced systematics.

\begin{figure}[!htb]
  \centering
   
   \includegraphics[width=3.5in]{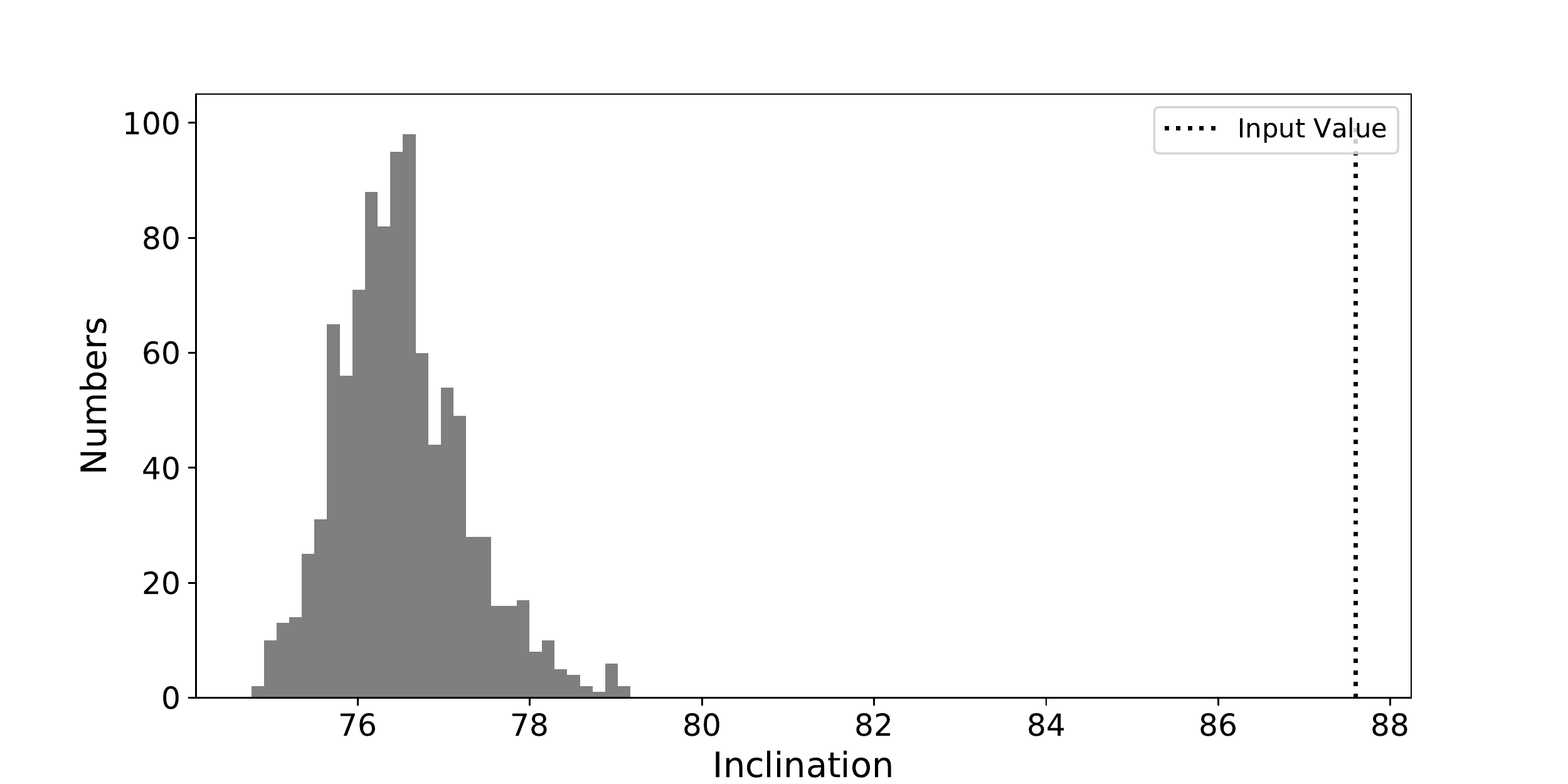}
   \includegraphics[width=3.5in]{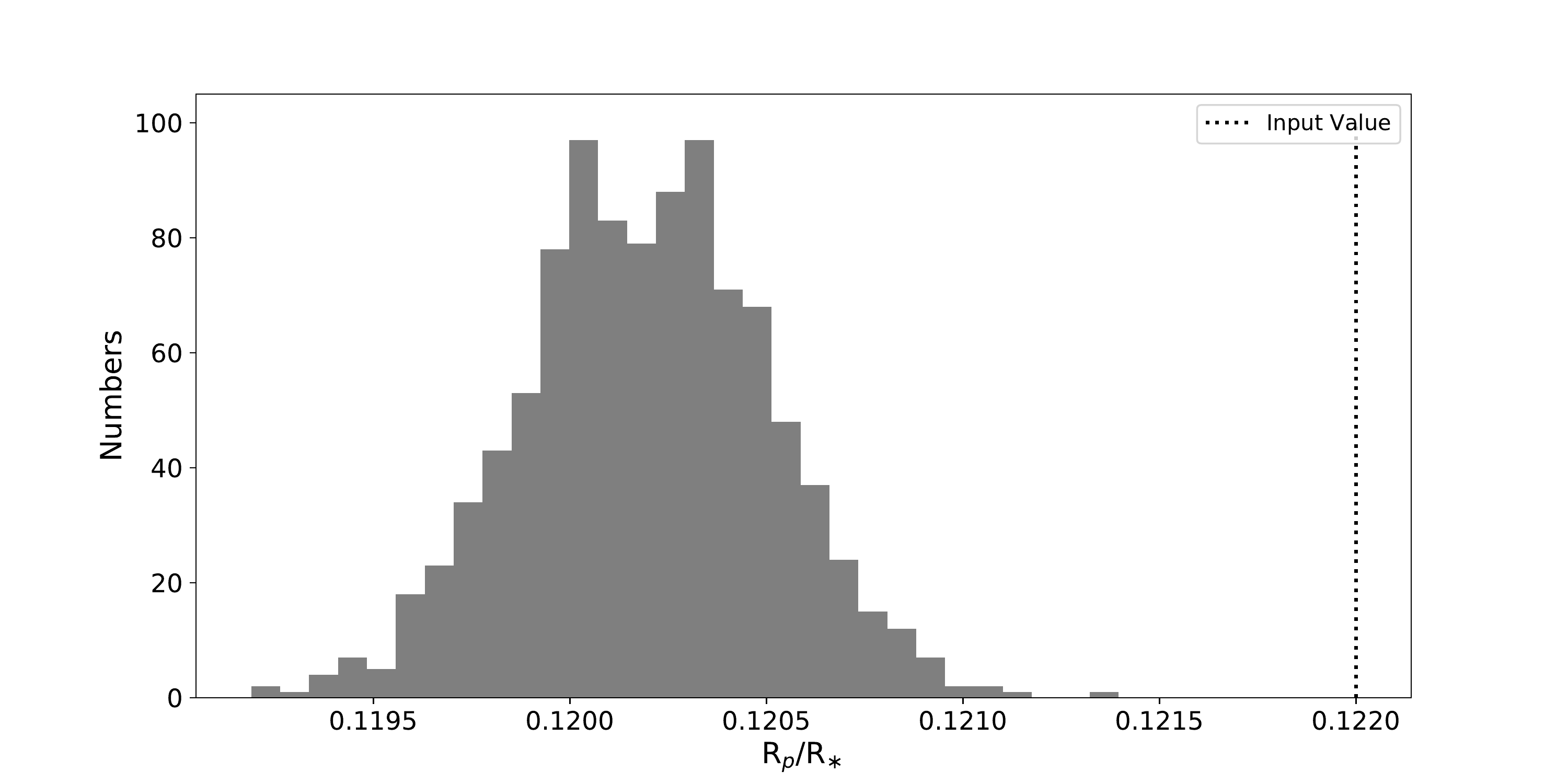}
  \caption{Top: The derived inclination distribution from fitting 1000 simulated light curves sampled at 30 minutes, for the parameters of WASP-121b. Bottom: The planet to star radius ratio distribution when fixing the inclination and semi-major axis. The vertical lines for both panels give the input values to the simulations. Clearly, for WASP-121b, a light curve cadence of 30 minutes would result in a large bias in derived parameters.}
  
\end{figure}

In the case of light curves with 2 minute sampling, the bias is assessed in the same way. The best fit inclination for WASP-121b is found to be 87.7$\pm$1.3$\deg$. The planet to star radius ratio with a fixed inclination and semi-major axis is 0.1219$\pm$0.0004. We find the difference between the derived values and the input values to be negligible with the 2 minute data.

For all three sources, we find that
the simulations with 2 minute sampling yield similar parameters as the input value. Meanwhile, with 30 minutes cadence, the result is significantly biased when we do not correct for the undersampling \citep{David2010,Kreidberg2015}. Also, the differences in the inclination and planet-to-star radius ratio between 2 and 30 minutes cadence matches the difference between them from real TESS data. From these simulation, we therefore conclude that the biases in parameters from fitting TESS data of different cadence comes from binning, and that the fit parameters from the 2 minute cadence data for our three targets are robust.

Finally, we undertake a simulation to investigate the dependence of inclination and semi-major axis upon sampling time interval when not applying the oversampling technique (Figure 8). The sampling time interval in the simulation ranges from 1 minute to 50 minutes. For each interval, the light curve is generated as described above and fit. We duplicate 10 simulations for each sampling interval. The median fit values and standard deviations of the derived parameters are shown in Figure 8 for each sampling interval. It is clear that the inclination and semi-major axis are not significantly biased when the sampling time is less than 5 minutes. Also, the result indicates that the inclination and semi-major axis derived are distorted to be smaller when applying a larger sampling interval (as shown in Figure 8). This suggests that great care must be taken when comparing transit depths in different bands, especially when the light curve sampling intervals are likely to be quite different. The simulations were conducted without the oversampling correction \citep{David2010} which can mitigate the bias of inclination and semi-major axis as previously discussed.

\begin{figure}[!htb]
  \centering
    \includegraphics[width=3.5in]{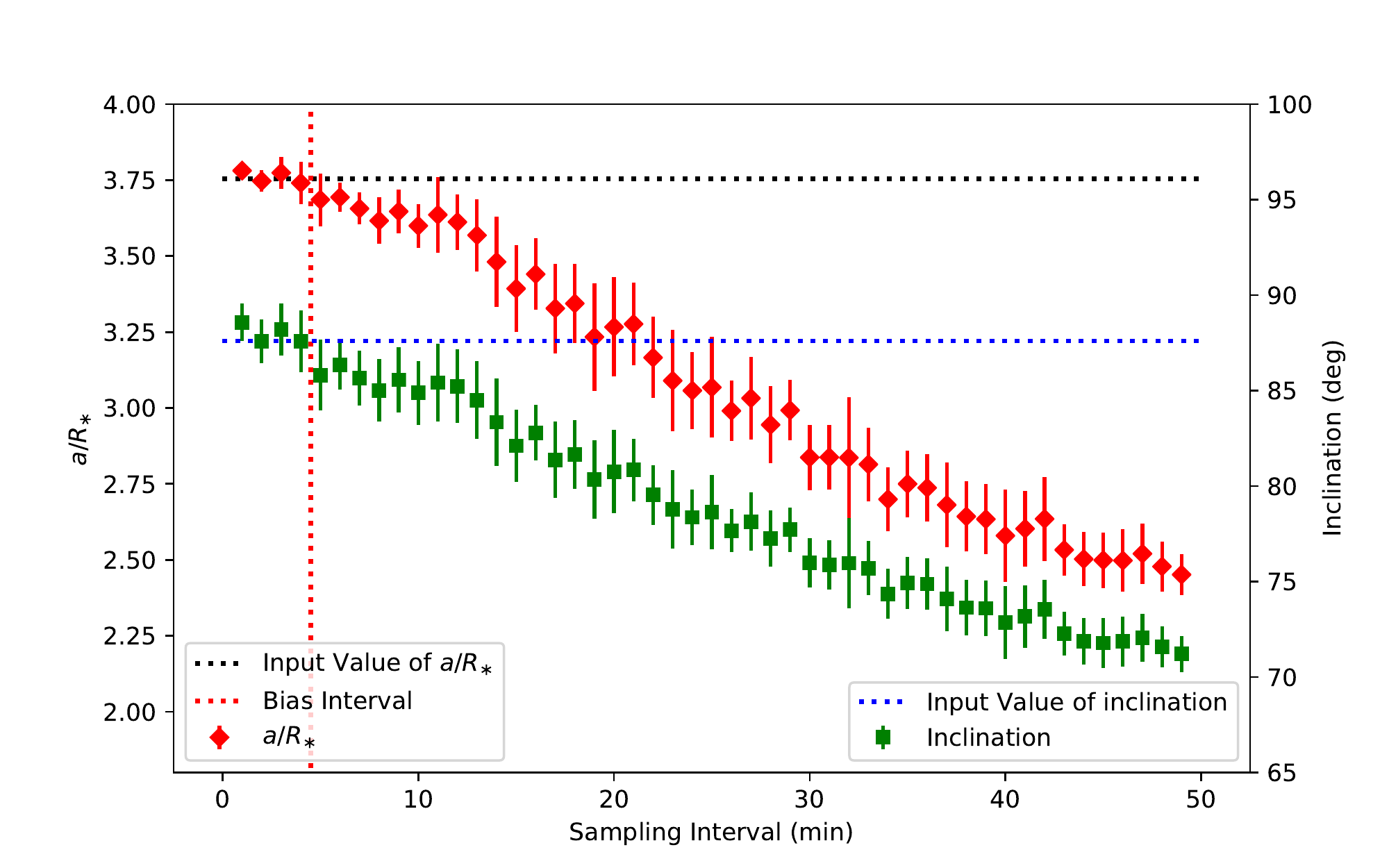}
  \caption{Results from fitting inclination (green squares) and semi-major axis (red diamonds) to simulated light curves of different sampling intervals when not correcting for the undersampling. The vertical red dotted line gives the sampling interval limit below which the parameters derived are not biased and corresponds to 5 minutes. The horizontal black line dotted is the input value of the semi-major axis; the dotted blue line is the input value of inclination. Clearly, TESS FFI 30 minute cadence data would result in significantly biased parameters for planets with short duration transits such as those considered here.}
\end{figure}

\begin{table*}
\setlength{\tabcolsep}{3mm}
\begin{center}
\caption{Planet Parameters}
\label{tab1}
\begin{tabular}{ccccc}
  \hline
  \hline
\small
{Parameters }&       {Description }&    KELT-19Ab          & WASP-156b          &  WASP-121b    \\
             &               &  68$\%$ Confidence  & 68$\%$ Confidence & 68$\%$ Confidence  \\
             
 \hline
  \multicolumn{5}{c} {Best Fitting Parameters} \\
  \hline
 R$_{p}$/R$_{\ast}$ &  Planet/star radius ratio in TESS  & 0.09576$\pm$ 0.00082 &  0.0670 $\pm$ 0.0012  &   0.1234$\pm$0.0004\\
 i [deg]   & Inclination   &   88.7$\pm0.7$      &   90.1$\pm1.4$  & 89.9$\pm1.6$ \\
 a/R$_{\ast}$       & Semi-major axis of planet orbit in unit of stellar radii &    9.04$\pm0.19$  & 12.98$\pm+0.73$  &  3.81$\pm0.03$    \\
std of residual &  Parts per million  & 1253 & 2130 & 1763 \\ 
reduced $ \chi^{2}$ &  Reduced chi-square  &  1.001 &  1.0009  &  1.0004  \\
 \hline
  \multicolumn{5}{c} {Fitting Results with Fixed Inclination and Semi-major Axis} \\
  \hline
R$_{p}$/R$_{\ast}$ & Planet/star radius ratio in TESS& 0.09940$\pm0.00048$  &    0.0662$\pm0.0009$  &  0.1230$\pm0.00025$\\ 
                           & Planet/star radius ratio in other work & 
 0.10713$\pm$0.00092$^{a}$  &   0.0685$^{+0.0012}_{-0.0008}$$^{e}$    &  0.1220$^{+0.0005}_{-0.0005}$ $^{b,c,d}$ \\ 
difference significance   &  Difference between TESS and former work in $\sigma$ & 7.549   & 1.533   &  1.789 \\             
 
i [deg]     & Inclination   &  85.41$^{+0.34}_{-0.31}$$^{a}$  &  89.1$^{+0.6}_{-0.9}$$^{d}$  &  87.6$\pm$0.6$^{b,c}$ \\     
&&&&89.1$\pm$0.5$^{d}$ \\
a/R$_{\ast}$       & Semi-major axis of planet orbit in unit of stellar radii & 7.50$^{+0.20}_{-0.18}$$^{a}$ & 12.8$^{+0.3}_{-0.7}$$^{e}$  &  3.754$^{+0.023}_{-0.028}$$^{b,c}$ \\
&&&&3.86$\pm$0.02$^{d}$\\
std of residual  &  Parts per million  &  1266  &  2132  &  1765  \\
reduced $\chi^{2}$ &  Reduced chi-square  &  1.021  &  1.001  &  1.002  \\
linLimb            &  Linear  limb-darkening  coefficient     & 0.29$\pm$0.04  & 0.37 $\pm$0.04 & 0.24 $\pm$0.03 \\
quadLimb           &  Quadratic  limb-darkening  coefficient  & 0.23$\pm$0.04  & 0.23 $\pm$0.04 &  0.21 $\pm$0.04 \\

\hline
\end{tabular}
\end{center}
\begin{flushleft}
Reference: 

(a) \citet{KELT-19};

(b) \citet{Delrez};

(c) \citet{Evans};

(d) \citet{Evans2018};

(e) \citet{Demangeon};

\end{flushleft}
\end{table*}

\subsection{A difference in stellar activity level?}

As explained in \citet{pont2008, Agol2010}, differences in the number of starspots between past observations and the TESS observations could account for differences in the measured transit depth. It was estimated that when the flux density of the star changes by 1\% because of stellar spots, the influence on the radius ratio is of the order of 10$^{-3}$R$_{p}$/R$_{\ast}$, comparable to the uncertainty of our TESS derived value. None of the three sources studied in this paper \citep{KELT-19,Demangeon,Evans} have shown 
any past evidence of stellar activity. Furthermore, if starspots were significant, the standard deviation of the light
curve outside the transit should be different from that during transit. So we conclude that the impact
of starspots on our transit depth measurement is negligible.

\section{Atmospheric Constraints}
Our primary scientific motivation for undertaking this work was to compare $R_{p}/R_{\ast}$ in the TESS band with the value from previous work to constrain the atmospheres of the exoplanets. In particular, the TESS bandpass is significantly redder than the {\it Kepler} band and encompasses the
0.95\,$\mu$m water feature. The presence of these features in the planets'
atmosphere would result in a larger value of $R_{p}/R_{\ast}$ in the TESS band. For instance, the $R_{p}/R_{\ast}$ of KELT-19Ab in TESS data is significantly smaller, at the level of 7.5$\sigma$, indicating the possible existence of atmospheric features.

Our fixed inclination fits (Table 1) allow for a more direct comparison with past work. 
The significance level of differences in $R_{p}/R_{\ast}$ are 7.5, 1.5, 1.8 $\sigma$ for KELT-19Ab, WASP-156b and WASP-121b, respectively. We discuss the implications of this difference for each of the exoplanets.

\subsection{KELT-19Ab}

KELT-19Ab is a giant planet transiting a moderately bright (V $\sim$ 9.9\,mag) A8V star with a orbital period of 4.61 days \citep{KELT-19}. The host stars has $M_{\ast}=1.62^{+0.25}_{-0.20}M_{\odot}$, $R_{\ast}=1.83\pm0.10R_{\odot}$. The planet has a radius of $R_{P}=1.91\pm0.11R_{J}$.

The ground-based light curves from \citet{KELT-19} in each band ($B$, $g$, $r$, $i(7519\AA)$, $I(8317\AA)$ and $z(8992\AA)$) were obtained. They are re-fit, to compare the variation in $R_{p}/R_{\ast}$ values between the different bands to that presented for the combined band. The light curves are performed with binary component deblending and linear detrending, using the same method as described previously. The flux contamination fraction adopted is the same as \citet{KELT-19}. The ground-based light curve does not have a long and stable baseline as TESS data. The long-term trend is therefore very difficult to estimate and dominates the $R_{p}/R_{\ast}$ uncertainty.

We masked the transit phase in the light curve as described previously, and fit the baseline, using different baseline lengths. This is repeated 5 times with the baseline length reduced by a factor of 10$\%$ each time. Three detrended light curves with the best residual are used for transit fitting. The $R_{p}/R_{\ast}$ and uncertainty are taken as the median value and the standard deviation of three fitted values. We do not fit the $B$ and $r$-band data because of the short baseline and partial coverage of the transit. The $i-$band values from the different telescopes are combined. KELT-19Ab has $R_{p}/R_{\ast}$ of 0.10713 derived from a joint fit to ground-based (B, g, r, i, I, z) data. The refit of our result presents smaller values of $R_{p}/R_{\ast}$ in each band compared to what has previously been published 
\citep[0.10713,][]{KELT-19}. This indicates the $R_{p}/R_{\ast}$ differences might partially come from a different process choice.

We fitted the multi-band transit data with the grid of atmospheric transmission spectra from \citet{Goyal2019, Goyal2018} which are developed for hot, H$_{2}$/He dominated Jupiters.
This model is a 1D radiative-convective-equilibrium solution based on the ATMO code \citep{Tremblin2015,Tremblin2016, Drummond2016}. ATMO solves the radiative transfer for isothermal pressure-temperature (P-T) profiles, with opacity and chemical abundance being other parameters. The grid model has a parameter space of two chemical scenarios, 22 temperatures, 4 planetary gravities, 5 atmospheric metallicities, 4 C/O ratios, 4 scattering haze, 4 cloud parameters and a scaling factor to the specific planetary radius. Modifying specific chemical abundance will lead to different chemical equilibrium which is avoided in the use of the forward grid model.

In order to fit the model, we minimized the chi-square value between the observations and model predictions at all the bands. The model prediction at each band is derived by the following equation where $S_{\lambda}$ is the
filter response.

\begin{equation}
\frac{R_{p}}{R_{\ast}}=\frac{\int \frac{R_{p}}{R_{\ast}}_{\lambda}S_{\lambda}d{\lambda} }{\int S_{\lambda}d{\lambda}}     
\end{equation}

\begin{figure*}[!htb]
  \centering
\includegraphics[width=7in]{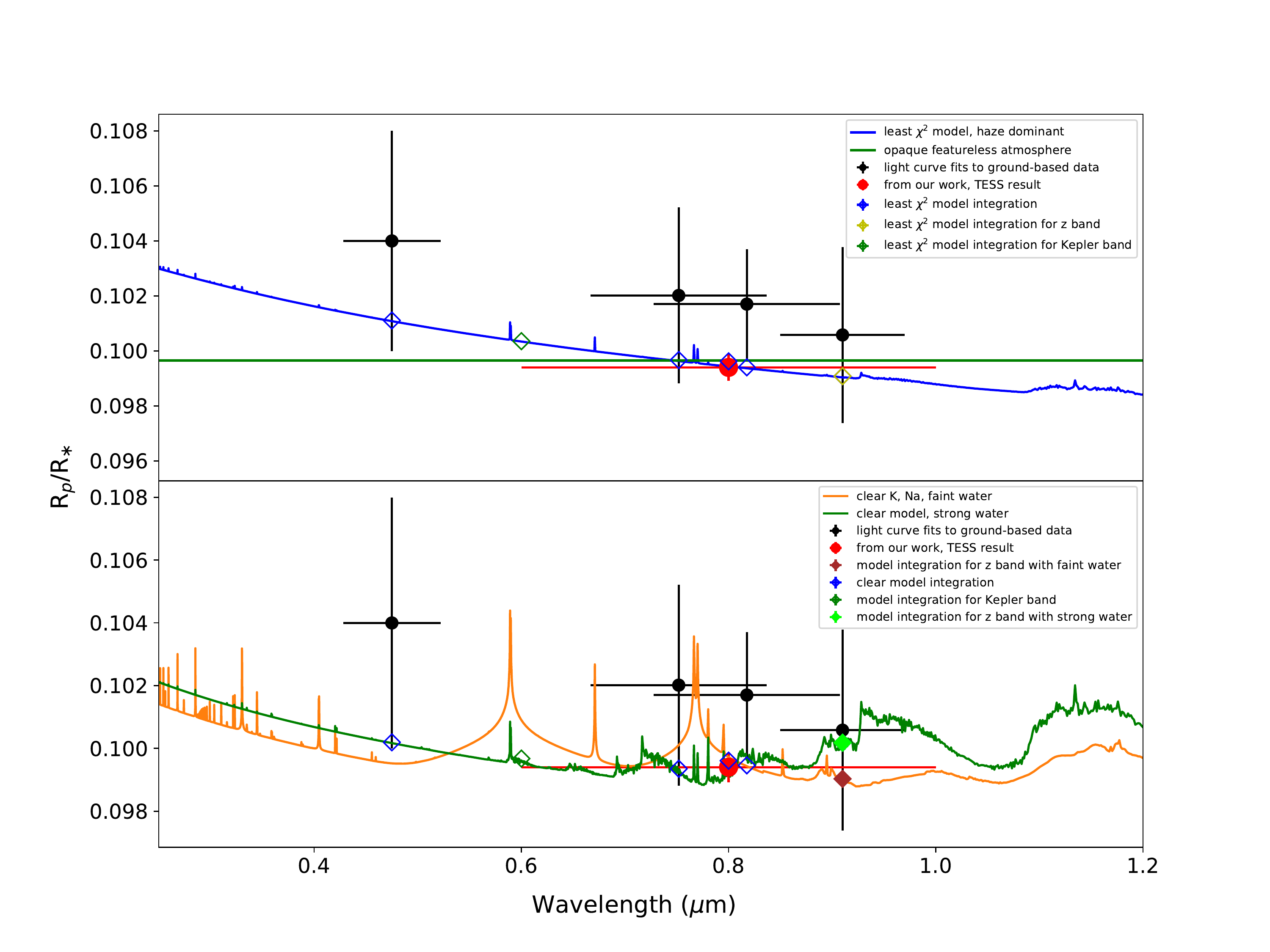}
  \caption{$R_{p}/R_{\ast}$ of KELT-19Ab in different bandpasses based on our re-analysis of those light curves. The red point is $R_{p}/R_{\ast}$ derived from TESS. The black points are $R_{p}/R_{\ast}$ from former work. In the top panel, the lowest $\chi$$^{2}$ model (haze dominant) is shown as a blue solid line. The green line indicates a flat, opaque featureless model that is consistent with the data. Blue diamonds are best-fit model predictions at the bands. The yellow diamond is the best fit model prediction at the SDSS $z$ band, green diamond is the best-fit model prediction at the Kepler band. In the bottom panel, a clear model with strong water (green) is shown with a faint water model (orange) as a comparison. Blue diamonds are strong water model predictions in different wavelengths bands while the green diamond is for the {\it Kepler} band. Dark green and brown diamonds are the strong water and faint water model prediction in the SDSS $z-$band. Clearly, 500ppm photometry in the $z-$band can distinguish between these scenarios.}
  \label{plot:keltindi}
\end{figure*}

We find that the haze dominated model at 1500K has the least $\chi$$^{2}$ of 2.80 (Figure \ref{plot:keltindi}). A clear model showing weak water vapor with rainout condensation at 1500K gives the second best $\chi^{2}$ of 3.05, while a clear model showing strong water vapor at 1000K has the third best $\chi^{2}$ of 3.26. A flat model with opaque featureless is also shown with a $\chi^{2}=3.14$. Other models resulted in much larger chi-square values (more than 6) and are not consistent with the data.  We also obtain the Bayesian Information Criterion (BIC) for evaluating the model evidence. BIC theory is effective when the data points are much more than the free parameters \citep{BIC}. This might not be the case for KELT-19Ab but we still present the BIC here for reference. The BIC is 4.41, 9.49, 9.69 for haze, weak water, and strong water models. The BIC is 4.74 for an opaque featureless model. 
The free parameter numbers of the opaque featureless model, haze, weak water, and strong water models are 1, 1, 4, and 4 respectively. The smaller $\chi^{2}$ and BIC of the haze model suggest that it is possibly the best interpretation given the current data. However, given the relatively narrow range in $\chi^{2}$ values and
small difference in BIC, 
the opaque featureless model would also be a reasonable explanation.

Based on our fits, we predicted the possible $R_{p}/R_{\ast}$ value in the {\it Kepler} band and $z-$ band based on these templates (Figure \ref{plot:keltindi}). The {\it Kepler} band would show $R_{p}/R_{\ast}$ of 0.1003 for a haze dominated atmosphere and 0.0997 for a clear model. In the $z-$band, 
the presence of water would introduce a significant difference in the ratio. We estimate that a clear atmosphere model with strong water contribution would show $R_{p}/R_{\ast}$=0.1002. The value is smaller, 0.0990, from a clear model with faint water, as well as from a haze dominant model.

We find that a difference of 1.2$\%$ in $R_{p}/R_{\ast}$ arises between strong water contribution and faint water contribution models. 
If we scale from our TESS measurements, and assume that the signal to noise
ratio depends on the square root of the number of points in the transit, we find that with 500ppm precision for 200 data points around the transit event in the $z-$band, one can distinguish between these two scenarios. This is of course a minimum since additional systematic effects can affect the measurements.

Alternately, instead of using our derived values of $R_{p}/R_{\ast}$ in each of the optical bands, we can use the value of \citet{KELT-19} which is derived from a joint fit to all the bands.
We treated their joint value of $R_{p}/R_{\ast}$ as the value in each band. The uncertainty follows the equation:
$\sigma_{joint}=\sigma_{i}/\sqrt{N}$, according to the error propagation law. $\sigma_{joint}$ is the joint band uncertainty, N is the number of bandpasses, $\sigma_{i}$ is the uncertainty in the band $i$. We assume the uncertainty in each band is the same.

Fitting $R_{p}/R_{\ast}$, the $\chi^{2}$ values for the different atmospheric models discussed - haze, strong water, weak water, and opaque featureless model - are 47.52, 50.68, 53.24, and 55.07, respectively. The BIC is 49.47, 58.46, 61.02, and 57.02 for the same models, respectively. Other models resulted in much larger chi-square values (more than 60) and are also not consistent with the data. Again, we find that a haze dominant atmosphere is preferred, as before.

\subsection{WASP-156b}

The discovery of WASP-156b was reported in \citet{Demangeon}. It is a super-Neptune orbiting a K type star (V=11.6\,mag, $M_{\ast}=0.842\pm0.052M_{\odot}$, $R_{\ast}=0.76\pm0.03R_{\odot}$) with a period of 3.84 days. It has a mass of $0.128^{+0.010}_{-0.009}M_{J}$, a radius of $0.51^{+0.02}_{-0.02}R_{J}$. 

\begin{figure*}[!htb]
  \centering
\includegraphics[width=7in]{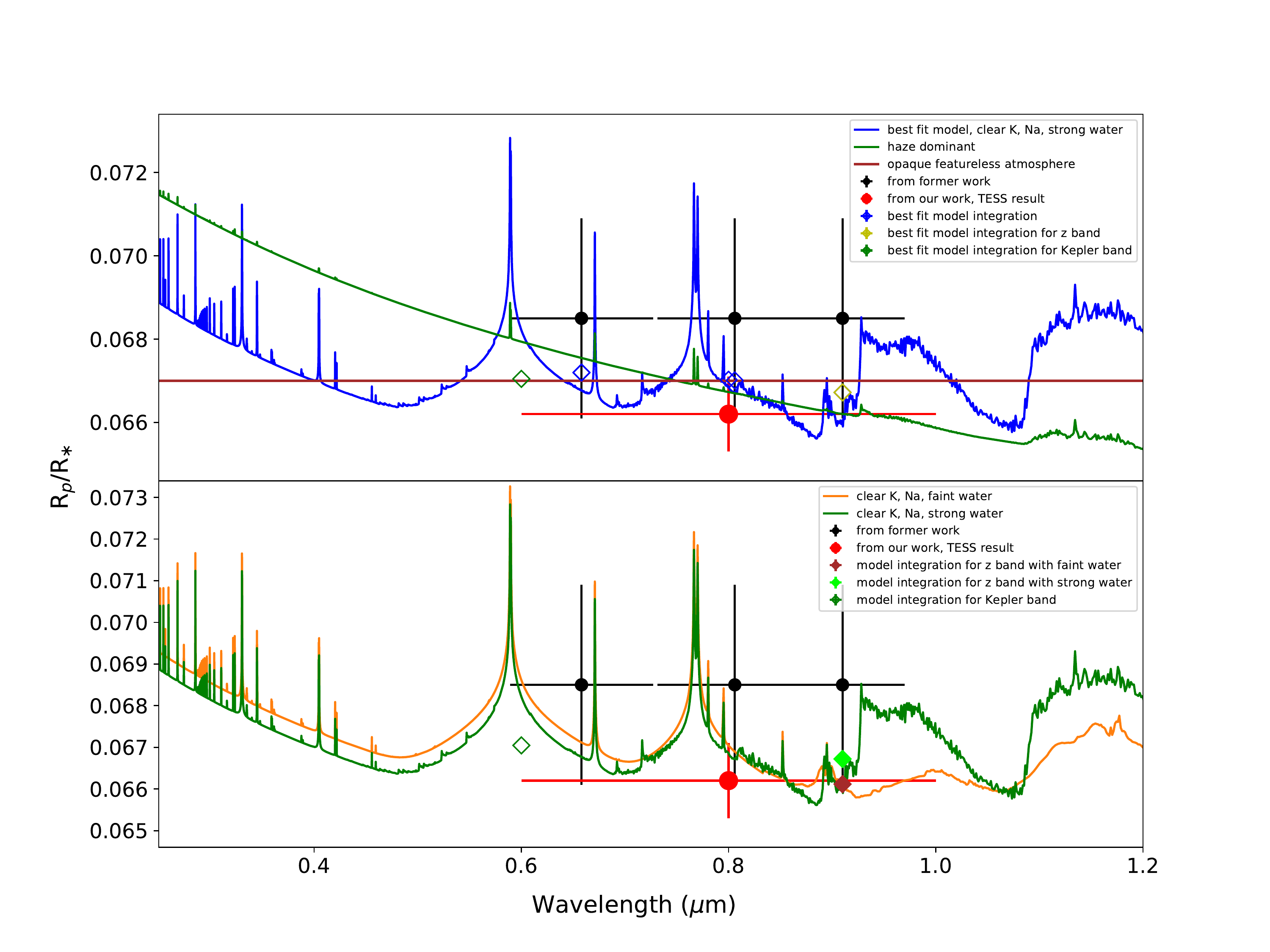}
  \caption{$R_{p}/R_{\ast}$ of WASP-156b in different bandpasses. The top panel shows the clear model with strong water (blue solid line), the haze dominant model  (green solid line) and the opaque featureless model (brown solid line).
  The red point is $R_{p}/R_{\ast}$ derived from TESS. The black points are $R_{p}/R_{\ast}$ from former work with the blue diamonds the "clear model" predictions at those bands. The yellow diamond is the model prediction at the SDSS $z-$band, green diamond is the model prediction at Kepler band. In the bottom panel, a clear model with strong water (green solid line) is shown along with a faint water model (Orange solid line) for comparison. 
  The empty green diamond is a strong water model prediction at Kepler band. The dark green diamond is the strong water model prediction at the SDSS $z-$band. Brown diamond is faint water model prediction at $z-$band.}
\label{plot:156}  
\end{figure*}

We followed the same procedure as for KELT-19Ab, to analyze the differences in $R_{p}/R_{\ast}$ in different bands. The published value of $R_{p}/R_{\ast}$ for comparison with our TESS analysis, is taken from \citet{Demangeon} at Johnson R, Gunn r, Johnson I and Gunn z (Figure \ref{plot:156}). 

We find that a clear atmosphere at 1500K with strong water contribution (best fit model) has a $\chi^{2}$ of 2.35, and a BIC of 8.80. The faint water model at the same temperature follows with a worse $\chi^{2}$ of $\sim$ 2.44, and a BIC of 8.87. The haze model has a $\chi^{2}$ of 2.44 and a BIC of 4.04. An opaque featureless model has the same $\chi^{2}$ of 2.35 as the best fit model. The smallest BIC value of 3.96 indicates that the data might prefer the opaque featureless atmosphere model. The strong water contribution causes a shift of 0.9$\%$ in $z-$band prediction of $R_{p}/R_{\ast}$ compared to the faint water model. In the Kepler band, the haze dominant model has a larger $R_{p}/R_{\ast}$ of 1.3$\%$ than the no atmosphere model. We conclude that the current measurements are consistent with an opaque featureless model.

\subsection{WASP-121b}
WASP-121b was discovered by \citet{Delrez}. It is a hot Jupiter around a main sequence star  (V=10.4\,mag, $1.353^{+0.080}_{-0.079}M_{\odot}$, $1.458\pm0.030R_{\odot}$) with a period of 1.28 days. The planets has a mass of $1.183^{+0.064}_{-0.062}M_{J}$ and a radius of $1.865\pm0.044R_{J}$.

\begin{figure*}[!htb]
  \centering
\includegraphics[width=7in]{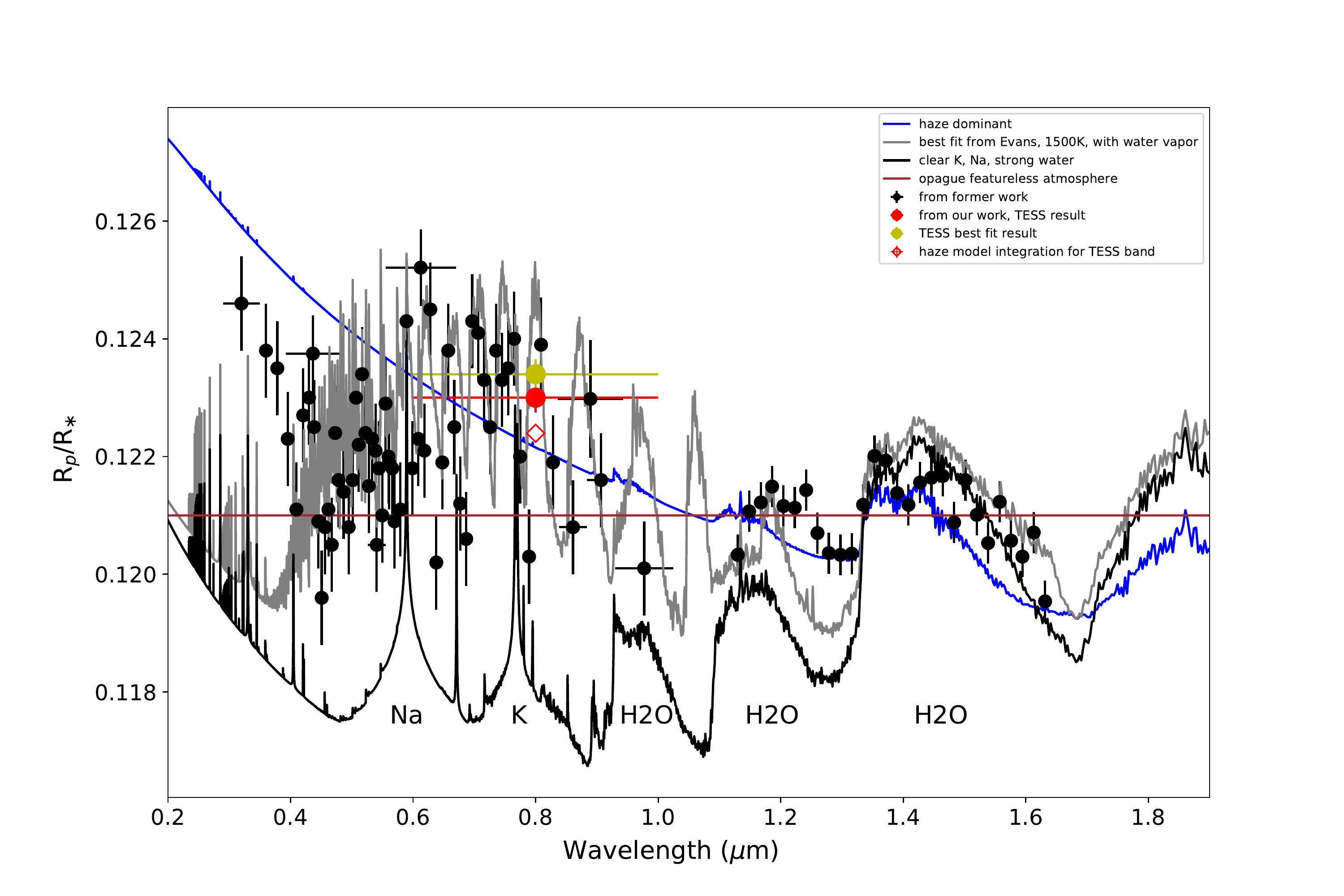}
  \caption{The transmission spectrum for WASP-121b. The solid yellow point is the best fit value of $R_{p}/R_{\ast}$ from the TESS light curve. However, if the inclination is fixed to previous work for a consistent comparison, the solid redpoint is $R_{p}/R_{\ast}$ derived from TESS data. The black points are $R_{p}/R_{\ast}$ from former measurements. 
  The grey solid line is the best fit from \citet{Evans2018}. The blue line shows the haze model. The brown line shows the opaque featureless model. 
  The black line indicates a clear atmosphere at a cool temperature (shifted for clarity). 
  \label{plot:121}
  }
\end{figure*}

We used external observations from \citet{Delrez, Evans2016, Evans2018}
in conjunction with our analysis of the TESS data (Figure \ref{plot:121}). 
These measurements utilized the standard ground-based filter, STIS, and WFC3 instruments on {\it Hubble} and span 0.3 to 1.65$\mu$m.
As outlined in \citet{Evans2018}, a global fit for the whole spectrum is challenging due to the complex problem of non-equilibrium chemistry.

Our fits with the inclination fixed to past work however, yield a value of $R_{p}/R_{\ast}$ which is in very good agreement with \citet{Evans2018}. The median value of the radius ratio is 0.1225 at the wavelength arrange of TESS. 
To fit the spectrum, we again use the general grid model from \citet{Goyal2019}.

The best-fit model from \citet{Evans2018} is still the best fit model when we add the TESS constraints. The fitting has a reduced chi-square of 4.91, a BIC of 449.74, with 92 data points, 4 fixed, and 4 free parameters (in Figure \ref{plot:121}). The fixed parameters are local condensation, atmospheric metallicities at 20x, haze parameters of 1.0, and no cloud. The fitted free parameters are temperature of 1500K, gravity at 10ms$^{-2}$, C/O ratio of 0.7, and planet radius (without atmosphere).
An opaque featureless model has a reduced chi-square of 5.19, a BIC of 477.39. The haze dominant model is with a chi-square of 5.87, a BIC of 538.76. A transmission model at 1500K with features of potassium, sodium, and water vapor has a much larger reduced chi-square of 27.40, a BIC of 2429.51, and seems to be ruled out. According to BIC theory, any model having a BIC smaller than 10 is favored \citep{BICmostcited}.

\section{SUMMARY AND DISCUSSION}
We have analyzed the TESS data for KELT-19Ab, WASP-156b, WASP-121b to 
measure $R_{p}/R_{\ast}$ in a broadband spanning 600nm to 1000nm. We identify the significant role of contamination by unresolved background stars in the TESS data and
use Gaia data to subtract their contribution to the transit light curves. 
We provide general-purpose software to address this issue for all TESS data.
We detrend the light curves through a linear fit to the photometry near the transit events. To derive the exoplanet transit parameters, we performed two ways of MCMC model fitting: with and without inclination and semi-major axis fixed. The former in particular helps with the comparison with past work on these stars. Using simulations, we demonstrate the systematic biases in inclination and radius ratio with 30 minutes cadence data when no correction for undersampling is undertaken \citep{David2010}. 
We also find the inclination and semi-major axis are distorted to be smaller with larger sampling times. However, 
we find that 2 minute sampling does not cause any bias for transit parameters. The atmosphere of these three exoplanets was then constrained based on $R_{p}/R_{\ast}$ in different bands.

In the case of KELT-19Ab, we find that the TESS-measured $R_{p}/R_{\ast}$ is $\sim$7.5$\sigma$ smaller than past work. The $R_{p}/R_{\ast}$ could be due to the choice of analysis or due to the presence of atmospheric features which further observations in narrower bands can reveal. The model fitting possibly favors a haze dominant atmosphere. In the case of WASP-156b, we find that the TESS transit depths agree with previous work and prefer a no atmosphere model fit to the measurements.
In the case of WASP-121b, the fits to the TESS light curve prefer the newest reported inclination. We find that the atmospheric transmission spectrum is best fitted with a 1500K model with water vapor as reported before.

Our work shows that exoplanet atmospheric model constraints benefit from high precision continuous photometry which provides strong constraints on key parameters such as inclination. 
Adding one high precision observation at certain bands improves the constraints on atmospheric composition. Since TESS
has now observed the Northern Hemisphere, 
we now have Kepler \citep{Kepler2010} and TESS constraints on a few bright stars, which will help assess the role of haze in exoplanet atmospheres. Finally, we show that precision photometry with $\sim$500ppm to 1000ppm in the $z-$band can help directly constrain the presence of water vapor in the exoplanet atmosphere, when combined with precise TESS photometry.

\acknowledgments
We are very grateful to the thoughtful comments by a referee which greatly improved the quality of results presented in this manuscript.
This work made use of PyAstronomy\footnote{https://github.com/sczesla/PyAstronomy} \citep{pya} and the NASA Exoplanet Archive \citep{ExoplanetArchive}. We would like to thank Jayesh M. Goyal for useful inputs and discussion and Caltech Optical Observatories for supporting the visit by the lead author. We also thank Hilke Schlichting for useful feedback on the manuscript. We sincerely thank Karen A. Collins for providing KELT-19Ab ground-based light curves. Fan Yang and Ji-Feng Liu acknowledge funding from National Natural Science Foundation of China (NSFC.11988101), National Key Research and Development Program of China (No.2016YFA0400800) and National Science Fund for Distinguished Young Scholars (No.11425313).


\bibliographystyle{aasjournal}
\bibliography{ref}

\begin{thebibliography}{}
\expandafter\ifx\csname natexlab\endcsname\relax\def\natexlab#1{#1}\fi
\providecommand{\url}[1]{\href{#1}{#1}}
\providecommand{\dodoi}[1]{doi:~\href{http://doi.org/#1}{\nolinkurl{#1}}}
\providecommand{\doeprint}[1]{\href{http://ascl.net/#1}{\nolinkurl{http://ascl.net/#1}}}
\providecommand{\doarXiv}[1]{\href{https://arxiv.org/abs/#1}{\nolinkurl{https://arxiv.org/abs/#1}}}

\bibitem[{{Agol} {et~al.}(2010){Agol}, {Cowan}, {Knutson}, {Deming}, {Steffen},
  {Henry}, \& {Charbonneau}}]{Agol2010}
{Agol}, E., {Cowan}, N.~B., {Knutson}, H.~A., {et~al.} 2010, \apj, 721, 1861,
  \dodoi{10.1088/0004-637X/721/2/1861}

\bibitem[{{Aigrain} {et~al.}(2015){Aigrain}, {Hodgkin}, {Irwin}, {Lewis}, \&
  {Roberts}}]{AGP}
{Aigrain}, S., {Hodgkin}, S.~T., {Irwin}, M.~J., {Lewis}, J.~R., \& {Roberts},
  S.~J. 2015, \mnras, 447, 2880, \dodoi{10.1093/mnras/stu2638}

\bibitem[{{Akeson} {et~al.}(2013){Akeson}, {Chen}, {Ciardi}, {Crane}, {Good},
  {Harbut}, {Jackson}, {Kane}, {Laity}, {Leifer}, {Lynn}, {McElroy}, {Papin},
  {Plavchan}, {Ram{\'\i}rez}, {Rey}, {von Braun}, {Wittman}, {Abajian}, {Ali},
  {Beichman}, {Beekley}, {Berriman}, {Berukoff}, {Bryden}, {Chan}, {Groom},
  {Lau}, {Payne}, {Regelson}, {Saucedo}, {Schmitz}, {Stauffer}, {Wyatt}, \&
  {Zhang}}]{ExoplanetArchive}
{Akeson}, R.~L., {Chen}, X., {Ciardi}, D., {et~al.} 2013, \pasp, 125, 989,
  \dodoi{10.1086/672273}

\bibitem[{{Berta} {et~al.}(2012){Berta}, {Charbonneau}, {D{\'e}sert},
  {Miller-Ricci Kempton}, {McCullough}, {Burke}, {Fortney}, {Irwin}, {Nutzman},
  \& {Homeier}}]{Berta2012}
{Berta}, Z.~K., {Charbonneau}, D., {D{\'e}sert}, J.-M., {et~al.} 2012, \apj,
  747, 35, \dodoi{10.1088/0004-637X/747/1/35}

\bibitem[{{Borucki} {et~al.}(2010){Borucki}, {Koch}, {Basri}, {Batalha},
  {Brown}, {Caldwell}, {Caldwell}, {Christensen-Dalsgaard}, {Cochran}, \&
  {DeVore}}]{Kepler2010}
{Borucki}, W.~J., {Koch}, D., {Basri}, G., {et~al.} 2010, Science, 327, 977,
  \dodoi{10.1126/science.1185402}

\bibitem[{{Claret}(2018)}]{TESSLD}
{Claret}, A. 2018, \aap, 618, A20, \dodoi{10.1051/0004-6361/201833060}

\bibitem[{{Czesla} {et~al.}(2019){Czesla}, {Schr{\"o}ter}, {Schneider},
  {Huber}, {Pfeifer}, {Andreasen}, \& {Zechmeister}}]{pya}
{Czesla}, S., {Schr{\"o}ter}, S., {Schneider}, C.~P., {et~al.} 2019, {PyA:
  Python astronomy-related packages}.
\newblock \doeprint{1906.010}

\bibitem[{{Delrez} {et~al.}(2016){Delrez}, {Santerne}, {Almenara}, {Anderson},
  {Collier-Cameron}, {D{\'{\i}}az}, {Gillon}, {Hellier}, {Jehin}, {Lendl},
  {Maxted}, {Neveu-VanMalle}, {Pepe}, {Pollacco}, {Queloz}, {S{\'e}gransan},
  {Smalley}, {Smith}, {Triaud}, {Udry}, {Van Grootel}, \& {West}}]{Delrez}
{Delrez}, L., {Santerne}, A., {Almenara}, J.-M., {et~al.} 2016, \mnras, 458,
  4025, \dodoi{10.1093/mnras/stw522}

\bibitem[{{Demangeon} {et~al.}(2018){Demangeon}, {Faedi}, {H{\'e}brard},
  {Brown}, {Barros}, {Doyle}, {Maxted}, {Cameron}, {Hay}, {Alikakos},
  {Anderson}, {Armstrong}, {Boumis}, {Bonomo}, {Bouchy}, {Delrez}, {Gillon},
  {Haswell}, {Hellier}, {Jehin}, {Kiefer}, {Lam}, {Lendl}, {Mancini},
  {McCormac}, {Norton}, {Osborn}, {Palle}, {Pepe}, {Pollacco}, {Prieto-Arranz},
  {Queloz}, {S{\'e}gransan}, {Smalley}, {Triaud}, {Udry}, {West}, \&
  {Wheatley}}]{Demangeon}
{Demangeon}, O.~D.~S., {Faedi}, F., {H{\'e}brard}, G., {et~al.} 2018, \aap,
  610, A63, \dodoi{10.1051/0004-6361/201731735}

\bibitem[{{Deming} {et~al.}(2013){Deming}, {Wilkins}, {McCullough}, {Burrows},
  {Fortney}, {Agol}, {Dobbs-Dixon}, {Madhusudhan}, {Crouzet}, \&
  {Desert}}]{Deming2013}
{Deming}, D., {Wilkins}, A., {McCullough}, P., {et~al.} 2013, \apj, 774, 95,
  \dodoi{10.1088/0004-637X/774/2/95}

\bibitem[{{Drummond} {et~al.}(2016){Drummond}, {Tremblin}, {Baraffe},
  {Amundsen}, {Mayne}, {Venot}, \& {Goyal}}]{Drummond2016}
{Drummond}, B., {Tremblin}, P., {Baraffe}, I., {et~al.} 2016, \aap, 594, A69,
  \dodoi{10.1051/0004-6361/201628799}

\bibitem[{{Esposito} {et~al.}(2019){Esposito}, {Armstrong}, {Gandolfi},
  {Adibekyan}, {Fridlund}, {Santos}, {Livingston}, {Delgado Mena}, {Fossati},
  {Lillo-Box}, {Barrag{\'a}n}, {Barrado}, {Cubillos}, {Cooke}, {Justesen},
  {Meru}, {D{\'\i}az}, {Dai}, {Nielsen}, {Persson}, {Wheatley}, {Hatzes}, {Van
  Eylen}, {Musso}, {Alonso}, {Beck}, {Barros}, {Bayliss}, {Bonomo}, {Bouchy},
  {Brown}, {Bryant}, {Cabrera}, {Cochran}, {Csizmadia}, {Deeg}, {Demangeon},
  {Deleuil}, {Dumusque}, {Eigm{\"u}ller}, {Endl}, {Erikson}, {Faedi},
  {Figueira}, {Fukui}, {Grziwa}, {Guenther}, {Hidalgo}, {Hjorth}, {Hirano},
  {Hojjatpanah}, {Knudstrup}, {Korth}, {Lam}, {de Leon}, {Lund}, {Luque},
  {Mathur}, {Monta{\~n}{\'e}s Rodr{\'\i}guez}, {Narita}, {Nespral}, {Niraula},
  {Nowak}, {Osborn}, {Pall{\'e}}, {P{\"a}tzold}, {Pollacco}, {Prieto-Arranz},
  {Rauer}, {Redfield}, {Ribas}, {Sousa}, {Smith}, {Tala-Pinto}, {Udry}, \&
  {Winn}}]{Esposito2019}
{Esposito}, M., {Armstrong}, D.~J., {Gandolfi}, D., {et~al.} 2019, \aap, 623,
  A165, \dodoi{10.1051/0004-6361/201834853}

\bibitem[{{Evans} {et~al.}(2016){Evans}, {Sing}, {Wakeford}, {Nikolov},
  {Ballester}, {Drummond}, {Kataria}, {Gibson}, {Amundsen}, \&
  {Spake}}]{Evans2016}
{Evans}, T.~M., {Sing}, D.~K., {Wakeford}, H.~R., {et~al.} 2016, \apjl, 822,
  L4, \dodoi{10.3847/2041-8205/822/1/L4}

\bibitem[{{Evans} {et~al.}(2017){Evans}, {Sing}, {Kataria}, {Goyal}, {Nikolov},
  {Wakeford}, {Deming}, {Marley}, {Amundsen}, {Ballester}, {Barstow},
  {Ben-Jaffel}, {Bourrier}, {Buchhave}, {Cohen}, {Ehrenreich}, {Garc{\'{\i}}a
  Mu{\~n}oz}, {Henry}, {Knutson}, {Lavvas}, {Lecavelier Des Etangs}, {Lewis},
  {L{\'o}pez-Morales}, {Mandell}, {Sanz-Forcada}, {Tremblin}, \&
  {Lupu}}]{Evans}
{Evans}, T.~M., {Sing}, D.~K., {Kataria}, T., {et~al.} 2017, \nat, 548, 58,
  \dodoi{10.1038/nature23266}

\bibitem[{{Evans} {et~al.}(2018){Evans}, {Sing}, {Goyal}, {Nikolov}, {Marley},
  {Zahnle}, {Henry}, {Barstow}, {Alam}, \& {Sanz-Forcada}}]{Evans2018}
{Evans}, T.~M., {Sing}, D.~K., {Goyal}, J.~M., {et~al.} 2018, \aj, 156, 283,
  \dodoi{10.3847/1538-3881/aaebff}

\bibitem[{{Foreman-Mackey}(2016)}]{corner2016}
{Foreman-Mackey}, D. 2016, The Journal of Open Source Software, 1, 24,
  \dodoi{10.21105/joss.00024}

\bibitem[{{Gaia Collaboration} {et~al.}(2016){Gaia Collaboration}, {Brown},
  {Vallenari}, {Prusti}, {de Bruijne}, {Mignard}, {Drimmel}, {Babusiaux},
  {Bailer-Jones}, \& {Bastian}}]{gaia2016}
{Gaia Collaboration}, {Brown}, A.~G.~A., {Vallenari}, A., {et~al.} 2016, \aap,
  595, A2, \dodoi{10.1051/0004-6361/201629512}

\bibitem[{{Gaia Collaboration} {et~al.}(2018){Gaia Collaboration}, {Brown},
  {Vallenari}, {Prusti}, {de Bruijne}, {Babusiaux}, {Bailer-Jones}, {Biermann},
  {Evans}, \& {Eyer}}]{Gaia2018}
---. 2018, \aap, 616, A1, \dodoi{10.1051/0004-6361/201833051}

\bibitem[{{Gibson} {et~al.}(2012){Gibson}, {Aigrain}, {Roberts}, {Evans},
  {Osborne}, \& {Pont}}]{GPdetrending}
{Gibson}, N.~P., {Aigrain}, S., {Roberts}, S., {et~al.} 2012, \mnras, 419,
  2683, \dodoi{10.1111/j.1365-2966.2011.19915.x}

\bibitem[{{Goyal} {et~al.}(2019){Goyal}, {Wakeford}, {Mayne}, {Lewis},
  {Drummond}, \& {Sing}}]{Goyal2019}
{Goyal}, J.~M., {Wakeford}, H.~R., {Mayne}, N.~J., {et~al.} 2019, \mnras, 482,
  4503, \dodoi{10.1093/mnras/sty3001}

\bibitem[{{Goyal} {et~al.}(2018){Goyal}, {Mayne}, {Sing}, {Drummond},
  {Tremblin}, {Amundsen}, {Evans}, {Carter}, {Spake}, \& {Baraffe}}]{Goyal2018}
{Goyal}, J.~M., {Mayne}, N., {Sing}, D.~K., {et~al.} 2018, \mnras, 474, 5158,
  \dodoi{10.1093/mnras/stx3015}

\bibitem[{{Hogg} \& {Foreman-Mackey}(2018)}]{Hogg2018}
{Hogg}, D.~W., \& {Foreman-Mackey}, D. 2018, \apjs, 236, 11,
  \dodoi{10.3847/1538-4365/aab76e}

\bibitem[{Kass \& Raftery(1995)}]{BICmostcited}
Kass, R.~E., \& Raftery, A.~E. 1995, Journal of the American Statistical
  Association, 90, 773, \dodoi{10.1080/01621459.1995.10476572}

\bibitem[{{Kipping}(2010)}]{David2010}
{Kipping}, D.~M. 2010, \mnras, 408, 1758,
  \dodoi{10.1111/j.1365-2966.2010.17242.x}

\bibitem[{{Kreidberg}(2015)}]{Kreidberg2015}
{Kreidberg}, L. 2015, \pasp, 127, 1161, \dodoi{10.1086/683602}

\bibitem[{{Kurucz}(1993)}]{Kurucz1993}
{Kurucz}, R.~L. 1993, {SYNTHE spectrum synthesis programs and line data}

\bibitem[{{Liddle}(2007)}]{BIC}
{Liddle}, A.~R. 2007, \mnras, 377, L74,
  \dodoi{10.1111/j.1745-3933.2007.00306.x}

\bibitem[{MacKay(2002)}]{MacKay2002}
MacKay, D. J.~C. 2002, Information Theory, Inference \& Learning Algorithms
  (New York, NY, USA: Cambridge University Press)

\bibitem[{{Madhusudhan} {et~al.}(2011){Madhusudhan}, {Harrington}, {Stevenson},
  {Nymeyer}, {Campo}, {Wheatley}, {Deming}, {Blecic}, {Hardy}, {Lust},
  {Anderson}, {Collier-Cameron}, {Britt}, {Bowman}, {Hebb}, {Hellier},
  {Maxted}, {Pollacco}, \& {West}}]{Madhusudhan2011}
{Madhusudhan}, N., {Harrington}, J., {Stevenson}, K.~B., {et~al.} 2011, \nat,
  469, 64, \dodoi{10.1038/nature09602}

\bibitem[{{Mandel} \& {Agol}(2002)}]{Mandel_Agol2002}
{Mandel}, K., \& {Agol}, E. 2002, \apj, 580, L171, \dodoi{10.1086/345520}

\bibitem[{Patil {et~al.}(2010)Patil, Huard, \& Fonnesbeck}]{pymc}
Patil, A., Huard, D., \& Fonnesbeck, C.~J. 2010, J. Stat. Softw, 1

\bibitem[{{Pont} {et~al.}(2008){Pont}, {Knutson}, {Gilliland}, {Moutou}, \&
  {Charbonneau}}]{pont2008}
{Pont}, F., {Knutson}, H., {Gilliland}, R.~L., {Moutou}, C., \& {Charbonneau},
  D. 2008, \mnras, 385, 109, \dodoi{10.1111/j.1365-2966.2008.12852.x}

\bibitem[{Pukelsheim(1994)}]{Friedrich1994}
Pukelsheim, F. 1994, The American Statistician, 48, 88,
  \dodoi{10.1080/00031305.1994.10476030}

\bibitem[{{Ricker} {et~al.}(2015{\natexlab{a}}){Ricker}, {Winn}, {Vanderspek},
  {Latham}, {Bakos}, {Bean}, {Berta-Thompson}, {Brown}, {Buchhave}, \&
  {Butler}}]{Ricker2015}
{Ricker}, G.~R., {Winn}, J.~N., {Vanderspek}, R., {et~al.} 2015{\natexlab{a}},
  Journal of Astronomical Telescopes, Instruments, and Systems, 1, 014003,
  \dodoi{10.1117/1.JATIS.1.1.014003}

\bibitem[{{Ricker} {et~al.}(2015{\natexlab{b}}){Ricker}, {Winn}, {Vanderspek},
  {Latham}, {Bakos}, {Bean}, {Berta-Thompson}, {Brown}, {Buchhave}, {Butler},
  {Butler}, {Chaplin}, {Charbonneau}, {Christensen-Dalsgaard}, {Clampin},
  {Deming}, {Doty}, {De Lee}, {Dressing}, {Dunham}, {Endl}, {Fressin}, {Ge},
  {Henning}, {Holman}, {Howard}, {Ida}, {Jenkins}, {Jernigan}, {Johnson},
  {Kaltenegger}, {Kawai}, {Kjeldsen}, {Laughlin}, {Levine}, {Lin}, {Lissauer},
  {MacQueen}, {Marcy}, {McCullough}, {Morton}, {Narita}, {Paegert}, {Palle},
  {Pepe}, {Pepper}, {Quirrenbach}, {Rinehart}, {Sasselov}, {Sato}, {Seager},
  {Sozzetti}, {Stassun}, {Sullivan}, {Szentgyorgyi}, {Torres}, {Udry}, \&
  {Villasenor}}]{2015Ricker}
---. 2015{\natexlab{b}}, Journal of Astronomical Telescopes, Instruments, and
  Systems, 1, 014003, \dodoi{10.1117/1.JATIS.1.1.014003}

\bibitem[{{Seager} \& {Deming}(2010)}]{seager2010}
{Seager}, S., \& {Deming}, D. 2010, \araa, 48, 631,
  \dodoi{10.1146/annurev-astro-081309-130837}

\bibitem[{{Sing} {et~al.}(2011){Sing}, {Pont}, {Aigrain}, {Charbonneau},
  {D{\'e}sert}, {Gibson}, {Gilliland}, {Hayek}, {Henry}, \&
  {Knutson}}]{sing2011}
{Sing}, D.~K., {Pont}, F., {Aigrain}, S., {et~al.} 2011, \mnras, 416, 1443,
  \dodoi{10.1111/j.1365-2966.2011.19142.x}

\bibitem[{{Sing} {et~al.}(2016){Sing}, {Fortney}, {Nikolov}, {Wakeford},
  {Kataria}, {Evans}, {Aigrain}, {Ballester}, {Burrows}, \&
  {Deming}}]{sing2016}
{Sing}, D.~K., {Fortney}, J.~J., {Nikolov}, N., {et~al.} 2016, \nat, 529, 59,
  \dodoi{10.1038/nature16068}

\bibitem[{{Siverd} {et~al.}(2018){Siverd}, {Collins}, {Zhou}, {Quinn}, {Gaudi},
  {Stassun}, {Johnson}, {Bieryla}, {Latham}, {Ciardi}, {Rodriguez}, {Penev},
  {Pinsonneault}, {Pepper}, {Eastman}, {Relles}, {Kielkopf}, {Gregorio},
  {Oberst}, {Aldi}, {Esquerdo}, {Calkins}, {Berlind}, {Dressing}, {Patel},
  {Stevens}, {Beatty}, {Lund}, {Labadie-Bartz}, {Kuhn}, {Col{\'o}n}, {James},
  {Yao}, {Johnson}, {Wright}, {McCrady}, {Wittenmyer}, {Johnson}, {Sliski},
  {Jensen}, {Cohen}, {McLeod}, {Penny}, {Joner}, {Stephens}, {Villanueva},
  {Zambelli}, {Stockdale}, {Evans}, {Tan}, {Curtis}, {Reed}, {Trueblood}, \&
  {Trueblood}}]{KELT-19}
{Siverd}, R.~J., {Collins}, K.~A., {Zhou}, G., {et~al.} 2018, \aj, 155, 35,
  \dodoi{10.3847/1538-3881/aa9e4d}

\bibitem[{{Smith} {et~al.}(2012){Smith}, {Stumpe}, {Van Cleve}, {Jenkins},
  {Barclay}, {Fanelli}, {Girouard}, {Kolodziejczak}, {McCauliff}, {Morris}, \&
  {Twicken}}]{PDC}
{Smith}, J.~C., {Stumpe}, M.~C., {Van Cleve}, J.~E., {et~al.} 2012, \pasp, 124,
  1000, \dodoi{10.1086/667697}

\bibitem[{{Tremblin} {et~al.}(2016){Tremblin}, {Amundsen}, {Chabrier},
  {Baraffe}, {Drummond}, {Hinkley}, {Mourier}, \& {Venot}}]{Tremblin2016}
{Tremblin}, P., {Amundsen}, D.~S., {Chabrier}, G., {et~al.} 2016, \apjl, 817,
  L19, \dodoi{10.3847/2041-8205/817/2/L19}

\bibitem[{{Tremblin} {et~al.}(2015){Tremblin}, {Amundsen}, {Mourier},
  {Baraffe}, {Chabrier}, {Drummond}, {Homeier}, \& {Venot}}]{Tremblin2015}
{Tremblin}, P., {Amundsen}, D.~S., {Mourier}, P., {et~al.} 2015, \apjl, 804,
  L17, \dodoi{10.1088/2041-8205/804/1/L17}

\bibitem[{{Vanderburg} \& {Johnson}(2014)}]{SFF}
{Vanderburg}, A., \& {Johnson}, J.~A. 2014, \pasp, 126, 948,
  \dodoi{10.1086/678764}

\bibitem[{{Vidal-Madjar} {et~al.}(2003){Vidal-Madjar}, {Lecavelier des Etangs},
  {D{\'e}sert}, {Ballester}, {Ferlet}, {H{\'e}brard}, \& {Mayor}}]{Vidal2003}
{Vidal-Madjar}, A., {Lecavelier des Etangs}, A., {D{\'e}sert}, J.~M., {et~al.}
  2003, \nat, 422, 143, \dodoi{10.1038/nature01448}

\bibitem[{{Yang} {et~al.}(2021){Yang}, {Long}, {Liu}, {Shan}, {Guo}, {Zhang},
  {Yi}, {Zheng}, \& {Zhao}}]{YangLD}
{Yang}, F., {Long}, R.~J., {Liu}, J.-f., {et~al.} 2021, \aj, 161, 294,
  \dodoi{10.3847/1538-3881/abf92f}

\end{thebibliography}
\end{document}